\documentclass[twocolumn]{aastex62}
\usepackage{amsmath,amsfonts,amssymb}
\usepackage{rotating}
\usepackage{hyperref}
\usepackage{tocloft}
\usepackage{enumitem}
\usepackage{ltablex}
\usepackage{capt-of}

\newcommand\aastex{AAS\TeX}

\graphicspath{{./}{figures/}}
\submitjournal{PASP}
\accepted{Nov. 19th 2018}
\shorttitle{\aastex\ iLocater Simulator \& Pipeline}
\shortauthors{Bechter et al.}

\begin{document}

\title{\large{Instrument Simulator and Data Reduction Pipeline for the iLocater Spectrograph}}

\correspondingauthor{Eric B. Bechter}
\email{ebechter@nd.edu}

\author[0000-0001-8725-8730]{Eric B. Bechter}
\affil{University of Notre Dame\\
225 Nieuwland Science Hall\\
Notre Dame, IN 46656, USA}

\author{Andrew J. Bechter}
\affil{University of Notre Dame\\
225 Nieuwland Science Hall\\
Notre Dame, IN 46656, USA}

\author{Justin R. Crepp}
\affil{University of Notre Dame\\
225 Nieuwland Science Hall\\
Notre Dame, IN 46656, USA}

\author{Jonathan Crass}
\affil{University of Notre Dame\\
225 Nieuwland Science Hall\\
Notre Dame, IN 46656, USA}

\author{David King}
\affil{University of Cambridge\\
Madingley Road\\
Cambridge, CB3 OHA, UK}

\begin{abstract}
iLocater is a near-infrared (NIR) radial velocity (RV) spectrograph that is being developed for the Large Binocular Telescope in Arizona. Unlike seeing limited designs, iLocater uses adaptive optics to inject starlight directly into a single mode fiber. This feature offers high spectral resolution while simultaneously maintaining a compact optical design. Although this approach shows promise to generate extremely precise RV measurements, it differs from conventional Doppler spectrographs, and therefore carries additional risk. To aid with the design of the instrument, we have developed a comprehensive simulator and data reduction pipeline. In this paper, we  describe the simulation code and quantify its performance in the context of understanding terms in a RV error budget. We find that the program has an intrinsic precision of $\sigma < 5$ cm/s, thereby justifying its use in a number of instrument trade studies. The code is written in \textsc{Matlab} and available for download on GitHub. 

\end{abstract}

\section{Introduction}
\label{sect:intro}

A new generation of Doppler radial velocity (RV) spectrographs aim to use adaptive optics (AO) to inject starlight directly into single-mode fibers (SMF) \citep{Crepp_2016}. Unlike multi-mode fibers (MMF) or conventional slit-fed designs, SMFs propagate only the fundamental spatial mode, resulting in a stable Gaussian beam that is completely decoupled from input imaging conditions and fiber stresses \citep{Bland_Hawthorn_06, Bland_Hawthorn_10}. The small mode field of SMFs, typically only several microns in diameter in the near-infrared (by definition comparable to the wavelength of light), further allows for high spectral resolution observations while offering a compact opto-mechanical design \citep{schwab_14,crepp_14,Jovanovic_16}. Realizing the advantages of improved wavelength sampling and instrument stability would allow diffraction-limited spectrographs to reach instrumental noise floors well below 1~m/s \citep{BechterA_2018}. Moreover, ultra-high resolution measurements ($\mathrm{R\geq 150,000}$) present an opportunity to study asymmetries in the profiles of stellar absorption lines, a first step towards disentangling the effects of stellar variability (astrophysical ``jitter") from the signal of orbiting planets \citep{Davis_17}. 

The RV semi-amplitude induced by an Earth-mass planet orbiting in the habitable zone of a Sun-like star corresponds to a measurement precision of only $10^{-4}$ pixels. However, the majority of data reduction pipelines currently available have only been tested at the level of $\sim10^{-3}$ pixels. Due to the fact that many subtle effects are expected to be encountered at the sub-meter-per-second level, it is important to start developing data reduction pipelines early-on in the design process for new RV instruments \citep{fischer_16}. In the case of diffraction-limited systems, for which there is no archival repository of representative frames, numerical simulations can serve as an important alternative that captures the physics relevant to SMF spectrographs and its impact on RV extraction methods. Investing time into thorough models also provides the framework necessary for trade-studies that offer feedback during the design process. 

Modern optical design tools, e.g. Zemax OpticStudio or Synopsys \textsc{Code V}, can use the optical model to generate spectral traces and orders as they would be measured by a megapixel array \citep{gibson_16, Terrien_2014}. The optical footprint generated from these tools allows spectra to be mapped to a physical detector plane. However, additional details are needed to simulate the entire light path from stellar source through the atmosphere to the telescope and instrument. Specifically, this includes simulated spectral sources, such as stellar spectra or calibration sources, atmospheric emission and absorption spectra, wavelength dependent instrument efficiency curves, instrument point-spread-functions (PSFs), and the addition of photon noise, detector noise, and other sources of uncertainty. Self-consistently combining all of these tools into a powerful simulation program that, when paired with a data reduction pipeline, can be used to investigate a number of important topics before the instrument sees first light: 

\begin{enumerate}[nosep]

\item Inform instrument optical design (e.g. spectral order and intra-order spacing requirements);
\item Quantify RV error budget terms; 
\item Simulate entire RV surveys and analyze on-sky performance;
\item Investigate secondary science cases following commissioning;
\item Optimize calibration system parameters; and 
\item Verify algorithms used in the data reduction process.
\end{enumerate}
Such calculations are particularly valuable when connecting scientific goals of a program to hardware requirements through the instrument design review process and laboratory testing. 

Motivated by the benefits of detailed software modeling, we have developed an end-to-end simulator as well as an initial data reduction pipeline for the iLocater spectrograph: a diffraction limited PRV instrument that will be installed at the Large Binocular Telescope (LBT) in Arizona \citep{Crepp_2016}. In this paper, we present the iLocater instrument simulator (\S2), including calibration of spectral models, Doppler broadening and RV shifts (planets, barycentric motion, noise), atmospheric simulations, instrument throughput calculations, instrument response function simulations, and detector simulations. In \S3, we describe the structure of each part of the modular data pipeline: image processing, spectral extraction, wavelength calibration, pre-processing, and RV extraction. We demonstrate the simulator's intrinsic RV precision in \S4. Finally, we present a number of useful applications from simulations in \S5. The simulation code described has been designed to adaptable such that minor code modifications can introduce additional astronomical and calibration sources or even different spectrograph optical designs. In order to make these tools accessible for the astrophysics community, we have made the source code publicly available for download on GitHub.\footnote{\url{https://github.com/ebechter/InstrumentSimulator}} 
\begin{figure*}[p]
  \sbox0{\begin{tabular}{@{}cc@{}}
    \includegraphics[width=1.15\textwidth]{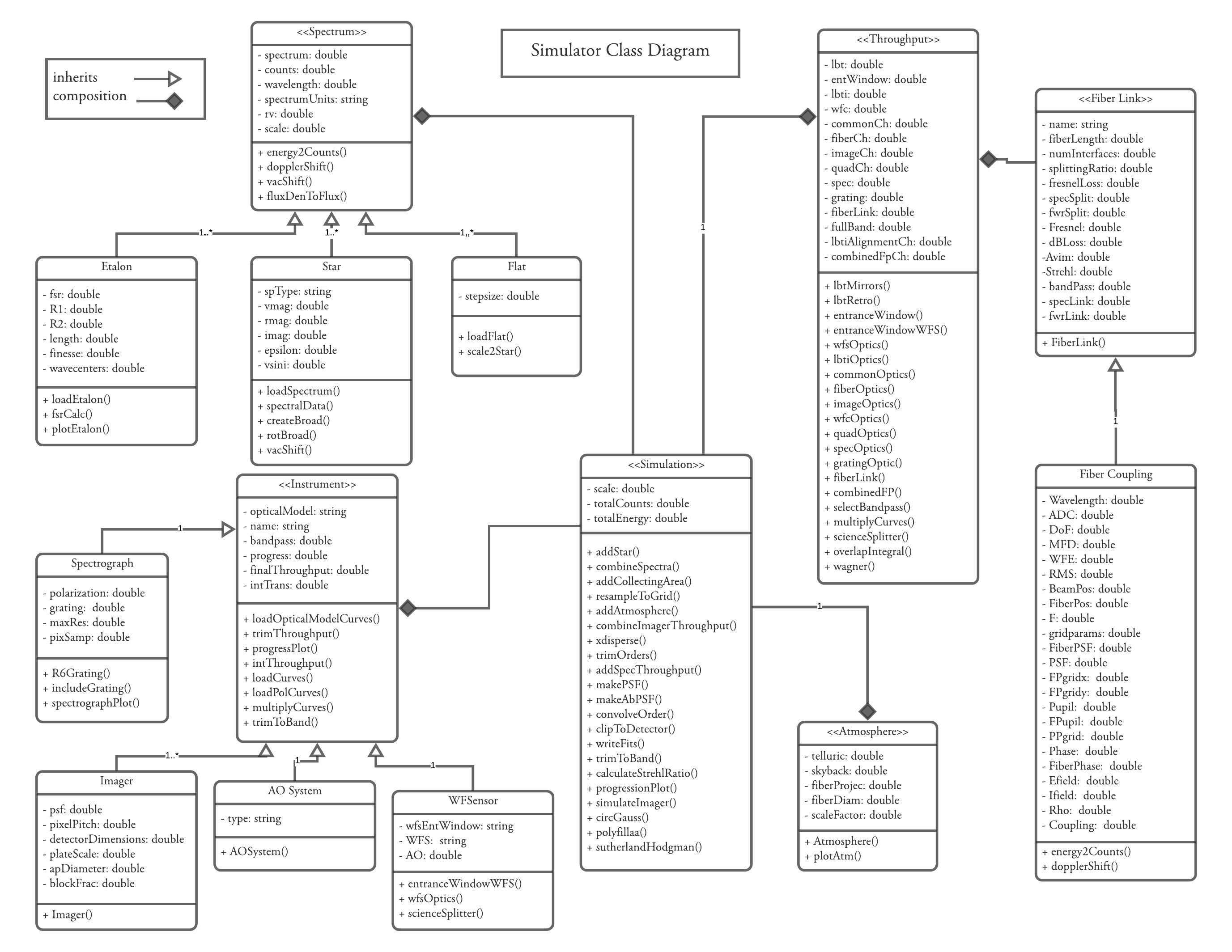}
  \end{tabular}}
  \rotatebox{90}{\begin{minipage}[c][\textwidth][c]{\wd0}
    \usebox0
    \captionof{figure}{UML class diagram for the Simulator. This chart shows the simulator's object-oriented nature for the six main super-classes and their sub-classes. Each super-class is named within $\mathrm{<<>>}$ and sub-classes are shown by the inherit open-tipped arrow. Composition lines note object existence dependencies and association line contains an indication of the one (1) or many (1..*) allowed instances of a class. Each box indicating a class show their properties first, including the data type and a list of their methods below.}
    \label{fig:classes}
  \end{minipage}}
\end{figure*}

\section{Instrument Simulator}\label{sec:simulator}
The software simulator uses an object-oriented, modular, class structure for its code architecture. The simulator has been developed initially for the iLocater system and is comparable to simulations built for instruments such as HPF and HARPS-N \citep{Terrien_2014, Artigau_2012}. The majority of code is written in \textsc{Matlab}, although it relies on optical design exports from Zemax, and detector noise generated from a Python package. The iLocater instrument is organized into three main parts: the acquisition camera, spectrograph, and calibration system \citep{Crepp_2016}. The acquisition camera couples AO-corrected light from the telescope into a single mode fiber that feeds the spectrograph. The calibration system provides the option of injecting a Fabry-P\'erot etalon spectrum, Uranium-Neon (U-Ne) emission lamp, or a white light spectrum, depending on the type of frame needed. Each of these optical systems are simulated in this code as well as the wavelength dependant throughput, AO system, atmosphere, and stellar spectra. 

\begin{figure*}[p]
		\includegraphics[width=0.95\textwidth]{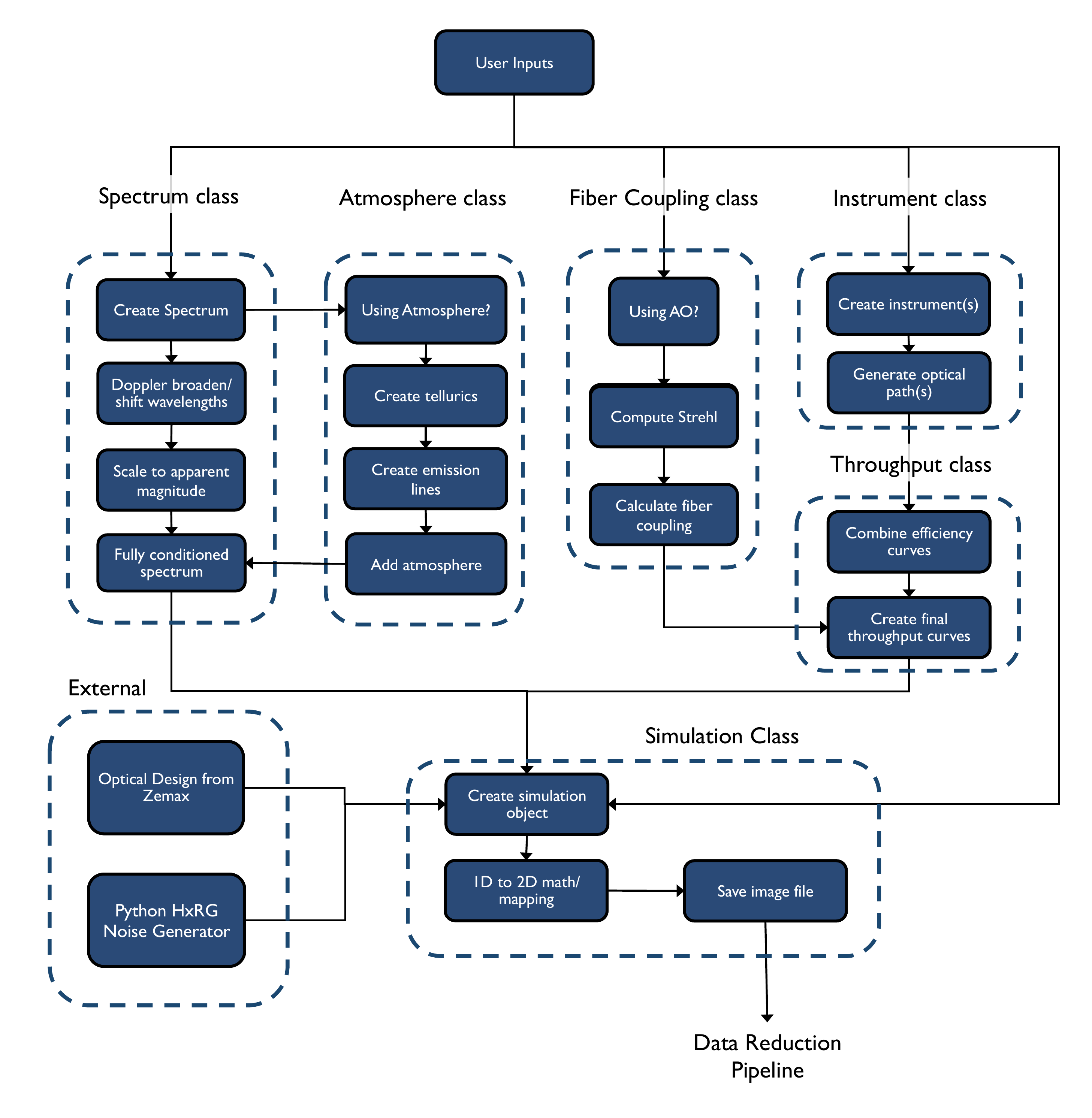} 
		\caption{Flow diagram of simulation code. Each class is responsible for generating the specific subclass of Spectrum or Instrument that will be used in the simulation class. Throughput curves are generated and condensed into a single efficiency curve which is combined with the final spectrum product in the simulation object. External noise generators and optical formatting files are also used by the simulation object.}
		\label{fig:codeflow}
\end{figure*}

The simulator class design uses the physical divisions of the iLocater instrument to define classes whenever possible (Figure~\ref{fig:classes}). There are six primary super-classes: (1) the Spectrum class instantiates the type of spectrum or spectra needed according to the mode selected by the user, (2) the Instrument class allows changes to the optical model to be made as well as modular selection of the instrument, (3) the Throughput class models each optical path of the instrument as a wavelength dependent efficiency curve, (4) the Simulation class combines results from the first two classes, imports Zemax optical design details, and handles converting spectra from 1D to 2D and binning them on the detector in a flux preserving manner, (5) the Atmosphere class creates atmospheric spectra according to specific input conditions, and (6) the Fiber Coupling class, which computes coupling losses due to fiber misalignments and other coupling issues. The following sections describe the flow of the simulation code which also shown visually in Figure~\ref{fig:codeflow}.

\newcommand\T{\rule{0pt}{2.8ex}}       
\newcommand\B{\rule[-1.4ex]{2pt}{0pt}} 
\begin{table*}
\caption{List of input variables and initialization parameters. For a more complete description of these, see the Getting Started section in the readme file included with the Github distribution.} 
\begin{tabularx}{0.99\textwidth}{l c c X}
\hline
\hline
Name & Default & Unit & \multicolumn{1}{c}{Description}\\
\hline
\multicolumn{4}{l}{\underline{Parallel Settings:}} \\
parflag & 0 & 1/0 & flag indicating the use of parallel processing. \\
numworkers & - & - & set number of parallel processing cores.\\ 
\multicolumn{4}{l}{\underline{Simulation Inputs:}}\\
scale & 1 & - & up-sampling factor used to minimize numerical errors during simulation.\\
dname & - & - & assign directory where files are saved.\\
fname & - & - & set file names for simulated frames. \\
\multicolumn{4}{l}{\underline{Stellar Inputs:}}\\
type & M0V & - & stellar spectral type model from catalog. \\
vmag & 11 & - & apparent visual magnitude of star. \\
vsini & 2 & km/s & stellar rotational velocity.\\
rv & 0 & m/s & inject a stellar radial velocity.\\
epsilon & 1& - & limb darkening parameter used in rotational broadening.\\
\multicolumn{4}{l}{\underline{Etalon Inputs:}}\\
FSR & 10 & - & free spectral range of Fabry-P\'erot. \\
R1 & 0.93 & - & reflectively of first mirror in Fabry-P\'erot.\\
R2 & 0.93 & - & reflectively of second mirror in Fabry-P\'erot.\\
finesse & - & - & specify finesse as an alternative to R1 and R2.\\
rv & 0 & m/s & inject a radial velocity.\\
\multicolumn{4}{l}{\underline{Spectrograph Inputs:}} \\
nOrders & 36 & - & number of spectral orders generated on detector.\\
tracenum & [1,2,3] & - & specify which of the three traces to be used.\\
footprint & 12.22 & - & optical layout version to be used in simulations.\\
waveSolution & 0 & 1/0 & new wavelength solution (1) or load previous (0).\\
throughput & - & - & cell array of strings indicating path of light.\\
polarization & [0, 0.5] & - & the degree of polarization and P-fraction of light.\\
\multicolumn{4}{l}{\underline{Fiber Coupling Inputs:}} \\
fiberpos & [$\frac{4}{7}$, $\frac{4}{7}$, 13] & $\mathrm{\mu m}$ & global fiber position offset in X, Y, and Z. \\
rndfiber & - & 1/0 & Gaussian random fiber position. \\
fibermas & 42 & mas & fiber diameter in milliarcseconds.\\ 
\multicolumn{4}{l}{\underline{Observation Conditions:}} \\
zdeg & 45 & degrees & zenith angle. \\
AO & SOUL & - & AO system. LBT's FLAO or SOUL AO system.\\ 
tellurics & 1 & 1/0 & include telluric spectrum (1) or ignore (0).\\
skyback & 1 & 1/0 & include sky-background spectrum (1) or ignore (0).\vspace{2pt}\\
\hline
\hline
\label{tab:inputs}
\end{tabularx}
\end{table*}

\subsection{Input Options and Variables}

The simulator is a flexible and complex code package containing many user input options and settings that need to be initialized. These include: parallel settings to enable parallel processing; simulation variables that allow up-sampling for more precise calculations; spectral options for stellar, atmospheric and calibration sources; spectrograph design and polarization settings; the number of spectral orders and sub-orders to be simulated; fiber coupling parameters; and observing conditions. Table \ref{tab:inputs} lists input options and variables currently available to the user. 

\subsection{Preparing the spectra}

An instance of each spectral class is created according to the specific settings defined by the user inputs. Each of these classes inherits a common set of properties and methods from the parent Spectrum class as well as having their own specific properties and methods (Figure~\ref{fig:classes}). 

\subsubsection{Stellar spectra} 
\begin{figure*}[htbp!]
\includegraphics[width=\textwidth]{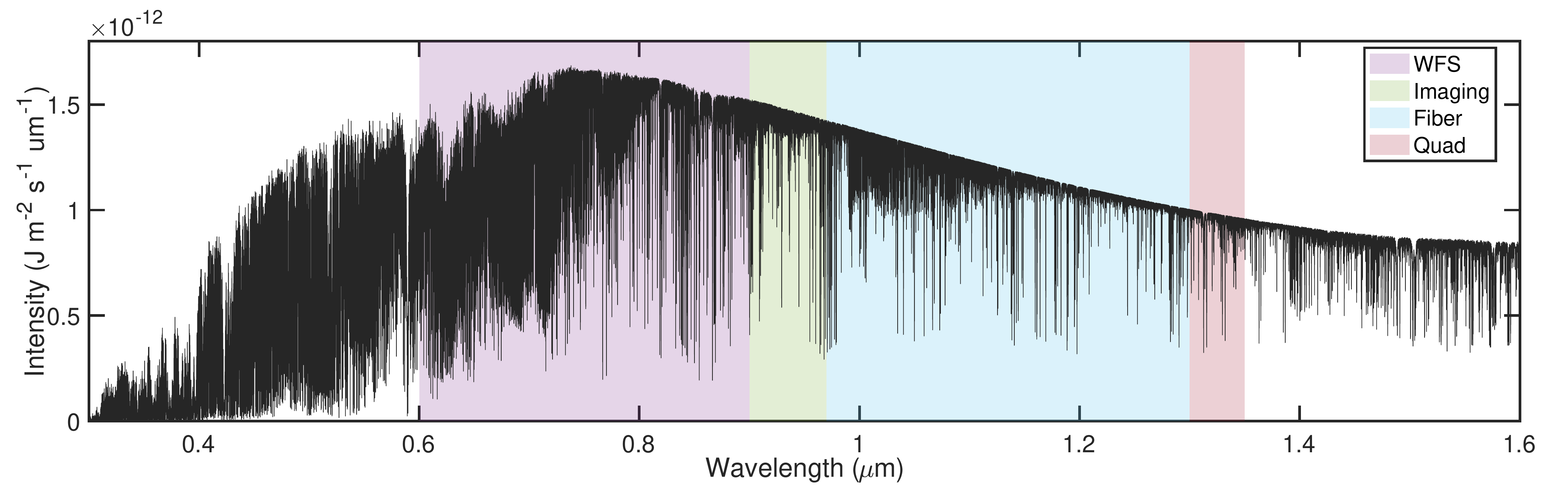}
\caption 
{\label{fig:inspec} BT-Settl synthetic spectrum: $\mathrm{T_{eff}=3600K}$ ($\sim$ M2V) scaled to an apparent magnitude of v=12. The shaded regions show the wavelengths of light used in each iLocater subsystem. The purple shaded band represents the wavefront sensing channel ($0.60-0.90 \mu m$), the green band is used in the acquisition camera imaging channel ($0.90-0.97 \mu m$), the blue band is directed to the the single-mode fiber spectroscopic channel ($0.97-1.27 \mu m$), and the red band is used in the quad cell ($1.3-1.5 \mu m$).} 
\end{figure*}

The simulator uses a catalog of BT-Settl FGKM models \citep{Allard_2012}. The requested spectrum is retrieved from the catalog and scaled to the specified apparent magnitude using the stellar color and effective temperature sequence table provided in \cite{Pecaut_13}. All spectra in the catalog are generated at vacuum wavelengths and a resolution of 0.02~\AA. Figure~\ref{fig:inspec} shows a model M2V spectrum, indicating iLocater's wavefront sensor, imaging, and spectroscopy bands.

Simulated stellar spectra first need to be Doppler broadened as it is largely responsible for the resulting profile of each line. Rotational line broadening manifests from stellar rotation, quantified by the projected equatorial velocity, $vsini$. Each surface element of the star with a different velocity will result in a different Doppler shift. To calculate the magnitude of this effect, the line-of-sight components of these surface velocities are integrated over the disk of the star, resulting in a cumulative Doppler shift distribution. To implement this rotational broadening effect in the simulator, the Doppler shift distribution is calculated according to Equation 18.14 in D. F. Gray's ``The Observation and Analysis of Stellar Photospheres'' \citep{Gray}.  This formulation also takes into account the relative flux levels changing from the stellar edge to the center via the linear-limb darkening law. 
Additionally, stellar spectra can be Doppler shifted according to a user input radial velocity. The relativistic Doppler equation is used to compute the wavelength shift:

\begin{equation}
\label{eq:Dopplershift}
\lambda_{shift} = \lambda * \sqrt\frac{1+\beta}{1-\beta} ,
\end{equation}
where $\lambda_{shift}$ is the Doppler shifted wavelength, $\lambda$ is the rest-frame wavelength, and $\beta=\Delta v_{rad}/c$. 

\subsubsection{Flat-field spectra}
There are two options for generating flat field spectra. The first source is  a discretely sampled uniformly flat spectrum that is scaled to a specified source power. This source is primarily used for debugging purposes and assessing throughput efficiences.  The second, more realistic, choice of flat-field source is a flattened white light spectrum generated from the same source that injects light into the Fabry-P\'erot calibration unit.

\subsection{Fabry-P\'erot Etalon}

Fabry-P\'erot etalons are a powerful and cost effective tool used for precise wavelength calibration in high resolution spectrographs. The etalon cavity produces a comb-like function in a way that can be tuned to match the the spectrograph wavelength band and resolution while offering numerous calibration peaks. In the simulator, the spectrum generated by a Fabry-P\'erot cavity is computed using the transmittance function: 
\begin{equation}
    T = \frac{1}{F \sin^2(\delta/2)},
\end{equation}
where F is the coefficient of finesse and delta is the phase difference between successive reflected pairs. Figure~\ref{fig:etalon} in \S\ref{dispersion solution} shows an example of a 10~GHz etalon. 
\S\ref{optimize etalon} outlines the optimization procedure used to select the specific parameters of iLocater's Fabry-P\'erot.

\subsection{Atmospheric Effects}

Accounting for the spectral contaminating effects of Earth's atmosphere (namely, tellurics and OH emission lines) is essential when designing an instrument to work in the NIR as telluric absorption features in this wavelength region can negatively impact RV measurements by several meters per second \citep{Bean_2010}. Telluric absorption features are comprised of a number of molecular species (e.g., H$_2$O, O$_2$, CH$_4$, CO$_2$) which contaminate stellar light as it travels through the atmosphere. Although few of these species have small seasonal variations (CO$_2$ \& CH$_4$), most will fluctuate on a much more frequent timescale (on the order of 10 minutes), presenting a significant obstacle to the data reduction process \citep{vacca_03}.

To simulate the absorption effects of Earth's atmosphere, a telluric spectrum is generated using the online TAPAS code which can then combined with stellar models \citep{TAPAS}. Given an RA, DEC, location (latitude, longitude, altitude), and a date, TAPAS computes the atmospheric transmittance from the top of the atmosphere down to the observatory, based on the HITRAN (high-resolution transmission molecular absorption database) molecular database and the LBLTRM (Line-By-Line Radiative Transfer Model) radiative transfer code. The program also provides separate transmittances associated with H2O, O2, O3, CO2, CH4, N2O and Rayleigh scattering. 

\begin{figure*}
\includegraphics[width=\textwidth]{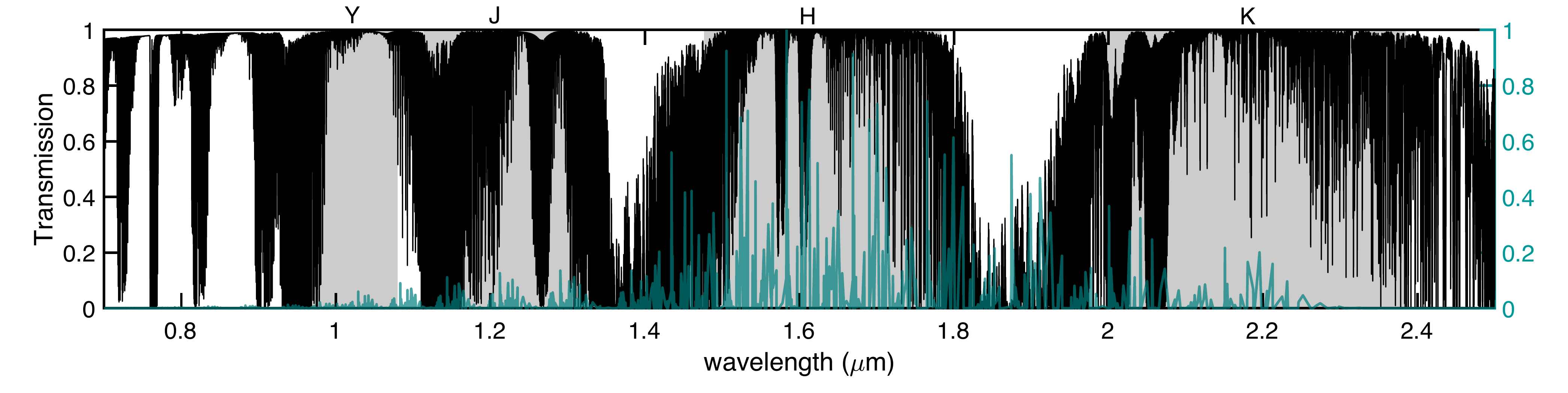}
\caption { \label{fig:tellurics} Transmission spectrum of the atmosphere in the NIR. The Y-, J-, H- and K-bands are shaded. Relative OH emission line strengths are shown in teal. iLocater has been designed to work in the Y- and J-band regions due to the relative lack of telluric and emission lines compared to other NIR regions.} 
\end{figure*}

Atmospheric emission lines at the NIR (1-2.5$\mathrm{\mu m}$) are mainly comprised of numerous narrow OH emission lines. These lines fluctuate on a timescale of 5-15 minutes with relative flux variations between $5$ and $10\%$. To incorporate sky background into the simulator, high resolution, simulated sky background spectra are downloaded from the Gemini website\footnote{http://www.gemini.edu/sciops/telescopes-and-sites/observing-condition-constraints/ir-background-spectra}. These models incorporate high resolution sky emission spectra, zodiacal continuum emission approximated by a T=5800~K black body, and thermal emission from the atmosphere treated as a black body with T=273~K, available at a variety of airmasses and water vapor column depths. Figure~\ref{fig:tellurics} is an example synthetic telluric spectrum generated at the LBT site (Mt. Graham, Arizona) with standard observing conditions and a normalized sky background spectrum over-plotted.  

Simulating ground-based observations also includes a wavelength shift through the atmosphere. This is accounted for by using the IAU standard conversion:
\begin{align}
\frac{\lambda_{vac}-\lambda_{air}}{\lambda_{air}} &=6.4328\times10^{-5}+\frac{2.94981\times10^{-2}}{146 - \sigma^2}\nonumber \\
 &\quad +\frac{2.5540\times10^{-4}}{41-\sigma^2}
\end{align}
\label{eq:AirVac}

where $\sigma=10^4/\lambda$, with $\lambda$ in angstroms.

\subsection{Flux simulation}
We estimate incident flux, $F({\lambda})$, based on the number of photons per second delivered to each of the instrument's optical channels using,
\begin{equation}\label{eqn:flux}
F(\lambda) = \int \: s_* (\lambda) \: a(\lambda) \: T(\lambda) \: d\lambda, 
\end{equation}
where $s_*(\lambda)$ is the stellar spectrum, $a(\lambda)$ represents the atmospheric effects, and $T(\lambda)$ is the instrument throughput. 

Steps taken to estimate flux are: 
\begin{itemize}
	\item Rotationally broaden stellar models $s_*$ using a projected rotational velocity, $v \sin{i}$.
	\item Scale flux to desired apparent magnitude (using Vmag as reference).
	\item Construct atmospheric spectrum, $a(\lambda)$, using synthetic telluric models for atmospheric absorption and sky-background for atmospheric emission. \item Multiply spectrum by the throughput curve of the channel, $T = \left( T_{\rm spec}, T_{\rm WFS}, T_{\rm image} \right)$
\end{itemize}
	
Flux values ($\mathrm{J s^{-1} m^{-2} \mu m}$) are converted to detector counts first by multiplying by the collecting area of LBT, calculated using a primary mirror diameter of $D=8.22$~m with a secondary obstruction of $d=0.91$~m (11\% diameter). To convert from Joules to photon counts, each discrete spectral bin is divided by the average energy of a photon in that bin. The product is multiplied by the width of each bin (in microns) and multiplied by the integration time. 

\subsection{Adaptive optics}
Calculating expected Strehl ratio is important when simulating AO-fed instrument performance as it is necessary for fiber coupling estimation. The baseline AO system performance for a given star is calculated using a look-up table that converts the simulated number of incident photons per second at the wavefront sensor pupil plane to a Strehl ratio at the center of the V, R, Z, Y, J, H, and K bands. The table was provided by the LBT AO group, which calculates Strehl ratio for a given zenith seeing \citep{pinna_16}. To approximate Strehl degradation with increasing airmass, we incorporate the zenith distance, $z$, analytically: 
\begin{equation}\label{eqn:airmass}
S(\lambda,\theta,z) = \exp(-|\log_{e}(S_0(\lambda,\theta))|\sec{z}),
\end{equation}
where $S_0(\lambda)$ represents the Strehl ratio at $\sec{z}=1$. The result is a changing chromatic throughput dependent on airmass. 

\subsection{Modeling instrument throughput}

Careful modeling of instrument throughput is essential for precision instruments, especially high resolution spectrographs, as they are typically photon starved. Accurate throughput budget models are used to inform many design choices including: estimating observational efficiency through exposure times, determining an instrument bandpass, and limiting stellar magnitudes for science cases. 

Instruments are generated as individual objects, as seen in Figure~\ref{fig:classes}. This allows for each instrument to be simulated individually or as a combination of subsystems. The result of throughput modelling is to generate the wavelength dependent instrument throughput $T(\lambda)$ for each object and the specified combination of all instrument objects. 

To create an instrument object, the user can define an arbitrary number of sequential optical elements. Each element can be defined with a surface coating name, number of surfaces, number of passes, angle of incidence, etc. Efficiency curves for each surface are drawn from a database of standard coatings and throughput is numerically calculated by sequentially multiplying the efficiency curves $T(\lambda)$ at each surface. If polarization is specified, then S and P curves of each optic are taken into account and the total throughput and degree of polarization are computed.

For iLocater, the user can specify a number predefined optical paths including those which represent the optical channels of the telescope, acquisition camera, spectrograph and calibration source. Each optical path is instantiated as an object and object parameters are populated both from initially defined parameters and the interaction between other classes during the simulation process. New optical paths can be defined and used interchangeably with existing ones to model instrument upgrades or as a stand-alone object for new instruments. 

The object-oriented nature of the simulation code allows the user to quickly identify the sub-components that introduce highest throughput losses. Design changes that alter optical coatings or the number optics can be assessed by quantifying the system performance both at the local subsystem and the global scale of the entire instrument. 

\subsection{Creating the simulation object}

The Simulation class first imports the necessary optical model values from Zemax. To accomplish this, we use a Zemax macro to export focal plane coordinates (x, y) and wavelength ($\mathrm{\lambda}$) data for several points per spectral order. Each order has three fiber traces that are physically offset to avoid cross-contamination. This data will be used to generate a wavelength solution and a physical mapping of each wavelength to the detector. The simulation code also has the ability to import PSF models at each desired wavelength. The purpose of a PSF model is to determine the instrument response function and incorporate the effects of optical diffraction. Zemax's physical optics propagation (POP) analysis supplies the appropriate PSF information. 

Once the optical model information has been generated in Zemax and read into \textsc{Matlab}, the wavelength solution is created by fitting either a low order polynomial to each spectral order or using a 2D model of Chebyshev polynomials that fits all spectral orders at the same time. The specific model is also used in the wavelength calibration module in the data reduction pipeline and is described in detail in \S\ref{sec:wavecal}. 

Next, final throughput curves are combined with fully conditioned spectra from the appropriate subclass of spectrum used to generate them. While the spectra are still 1D at this point, they are split into respective spectral orders in preparation for individual PSF convolution. Depending on the inputs provided and the mode specified, the simulator can use a single representative PSF for a spectral order or, using a custom convolution algorithm, spatially-varying PSFs across every order. The simulator also has the ability to inject optical aberrations in simulated PSFs, allowing for users to simulate realistic PSFs without the need for POP analysis in Zemax.
\begin{figure*}[p]
		\includegraphics[width=\textwidth]{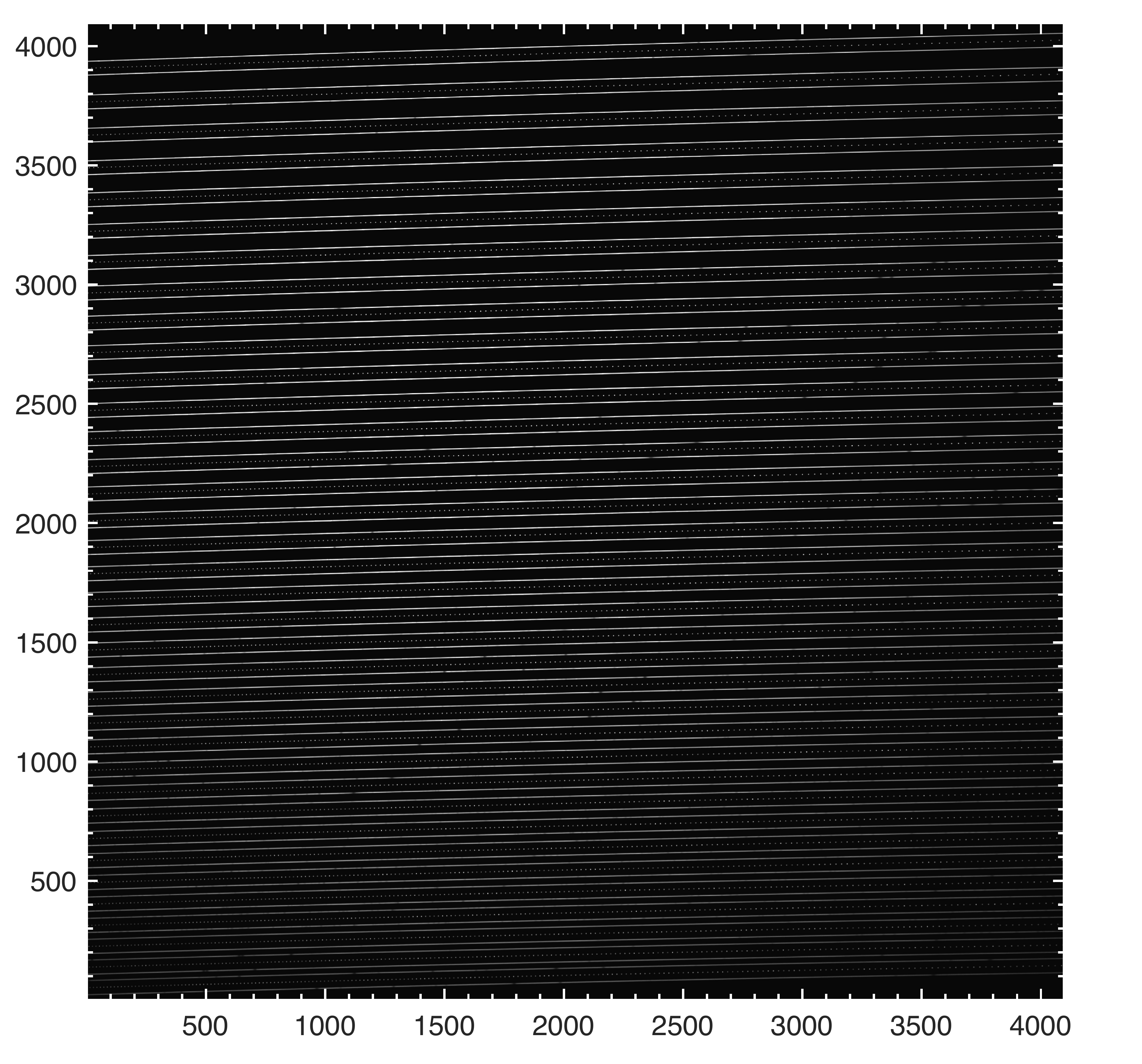} 
		\caption{Example of a simulated detector frame. The top and bottom traces in each spectral order contain the spectrum of an M0V star while the central trace displays simulated 10 GHz etalon spectrum for simultaneous calibration purposes. The 36 spectral orders span orders m = 117-152, top to bottom with the dispersion direction going left to right. There are no physical noise sources or detector effects shown in this frame, however the entire throughput of the telescope and instrument can be seen in the Gaussian shaped intensity profile across each order and along the cross-dispersion (vertical) direction.}
		\label{fig:fullframe}
\end{figure*}
After convolution, each spectral order has two spatial axes (x, y) and one wavelength axis. This rectangular array is then warped and clipped onto the detector sampling such that it matches the physical curvature and location of each order. Flux-preserving clipping is accomplished by assigning vertices to each discrete sample in the rectilinear array, representing each sample as a polygon. Each polygon's vertices are mapped to the detector coordinates using the previously acquired Zemax optical information and distributed into each pixel accordingly using a Sutherland-Hodgeman clipping algorithm.

Once the spectral signal has been computed across the detector, the user has the option of adding photon noise as well as detector noise. Photon-noise is governed by Poisson statistics and added to the array by a random draw from a Poisson distribution for each pixel. Detector noise is added by making use of a NIR detector system noise generator written for the James Webb Space Telescope Near Infrared Spectrograph \citep{Rauscher_2015}. Figure~\ref{fig:fullframe} shows a single output frame from the simulator with iLocater's spectral format imprinted on an H4RG detector. The array is saved as a fits file with a custom header that contains many of the specific parameter settings of the particular simulation run. 

\section{Data reduction pipeline}
\begin{figure*}[p]
	\begin{center}
		\includegraphics[width=0.95\textwidth]{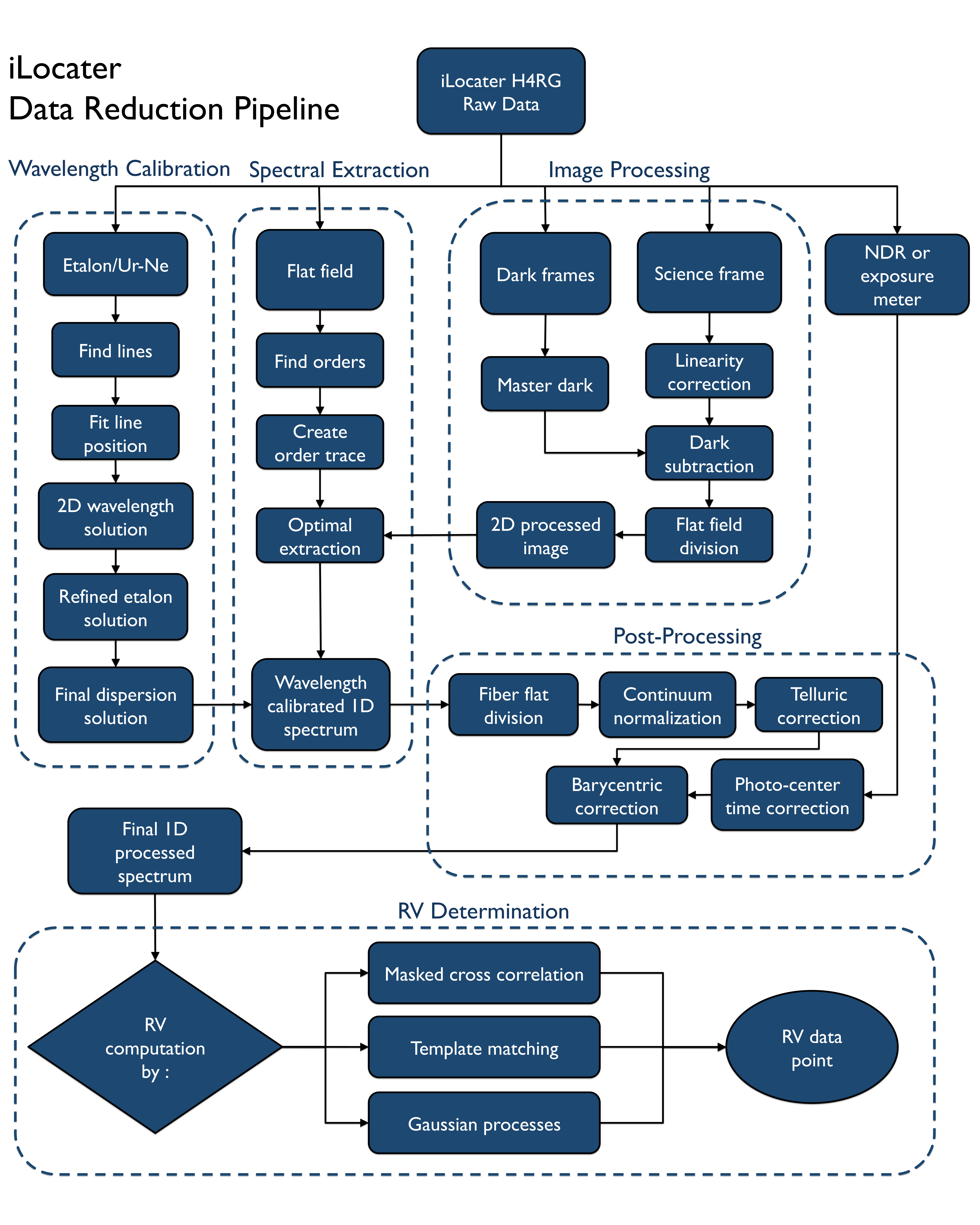}
		\caption{Data reduction pipeline organizational chart showing each sub-pipeline, their processes, and the flow of data required to recover a single RV measurement.}
		\label{fig:flow}
	\end{center}
\end{figure*}

The iLocater data reduction pipeline is comprised of five modular sub-pipelines: image processing, spectral extraction, wavelength calibration, post-processing, and RV computation. An overview of iLocater's data reduction pipeline is shown in Figure~\ref{fig:flow}, showing the flow of data required to recover a single RV measurement. All data reduction software algorithms have been developed in \textsc{Matlab}. The following section provides a description of each sub-pipeline.

\subsection{Image processing}

The image processing sub-pipeline converts raw science frames or cubes from the H4RG into cleaned, processed, 2D images, that are ready for spectral extraction. Figure \ref{fig:fullframe} shows an example raw image with iLocater's spectral format imprinted. The standard steps for our spectroscopic data reduction are:

\begin{enumerate}[noitemsep, nosep]
\item Master calibration frame creation, 
\item Pixel non-linearity correction,
\item Dark subtraction,
\item Bad pixel correction,
\item Flat division. 
\end{enumerate}

\subsubsection{Master calibrations}
The calibration frames used in image processing are flat fields and dark frames. Each master calibration frame is created using standard median combination. Flat fields are generated using the same white light source that powers the wavelength calibration unit combined with a flattening filter to minimize chromatic intensity modulations and neutral density filters to scale the output power according to the desired integration time.  

\subsubsection{Pixel nonlinearity}
HxRG's nonlinear response to light is a well documented effect in the literature, the most commonly mentioned effect being reciprocity failure \citep{Biesiadzinski_11, Plazas_17}. 
iLocater's H4RG non-linear response will be thoroughly tested and characterized in the laboratory, prior to spectrograph integration and on-sky commissioning. The image processing sub-pipeline will apply the prescribed linearizing factor for each pixel to restore count values in raw image frames. 

\subsubsection{Dark subtraction}
Dark frames are a way in which dark current, detector bias and other background sources can be subtracted from science and calibration images. Additionally, hot/dead or poorly responding pixels on the detector can be identified at this stage. Dark frames will be created by integrating for a time equal to the exposure time of science frames while blocking all light to the spectrograph or alternatively existing dark frames can be rescaled to match the exposure time of the science frame. As they will contain the same level of bias as other frames, separate bias frames are not necessarily needed. After linearity correction, master dark frames are subtracted from each frame.  
\subsection{Bad pixel correction}

Bad pixels identified in dark frames or during detector characterization in the laboratory are first saved in a bad pixel map, a 2D binary grid indicating the locations of bad pixels. Bad pixels are corrected by replacing each mapped bad pixel with a linearly interpolated value using a specified window surrounding the bad pixel.  

\subsubsection{Flat division}

Fiber-fed spectrographs, including iLocater, typically have no way to uniformly illuminate the entire detector once it is mounted in the spectrograph focal plane. Because of this they cannot correct full frame pixel-to-pixel variation, standard in most image processing pipelines. Fiber flats will be created by illuminating the entrance fibers with a white light source which will be used primarily to help remove the grating's blaze function (\S\ref{blazeremoval}). However, full frame flat-fields will be taken in the lab during detector characterization and testing. The fidelity of the lab flat-fields will be quantified throughout instrument commissioning.   

\subsection{Spectral extraction}

Spectral extraction is a particularly important step in the data reduction pipeline. In general, it refers to the operation of systematically recombining the signal around a central order's trace in the cross-dispersion direction, resulting in a 1D spectrum for each echelle order. 

\subsubsection{Order identification}

\begin{figure*}
	\begin{minipage}[t][][b]{.5\textwidth}
		\flushleft		
		\includegraphics[height=2.6in]{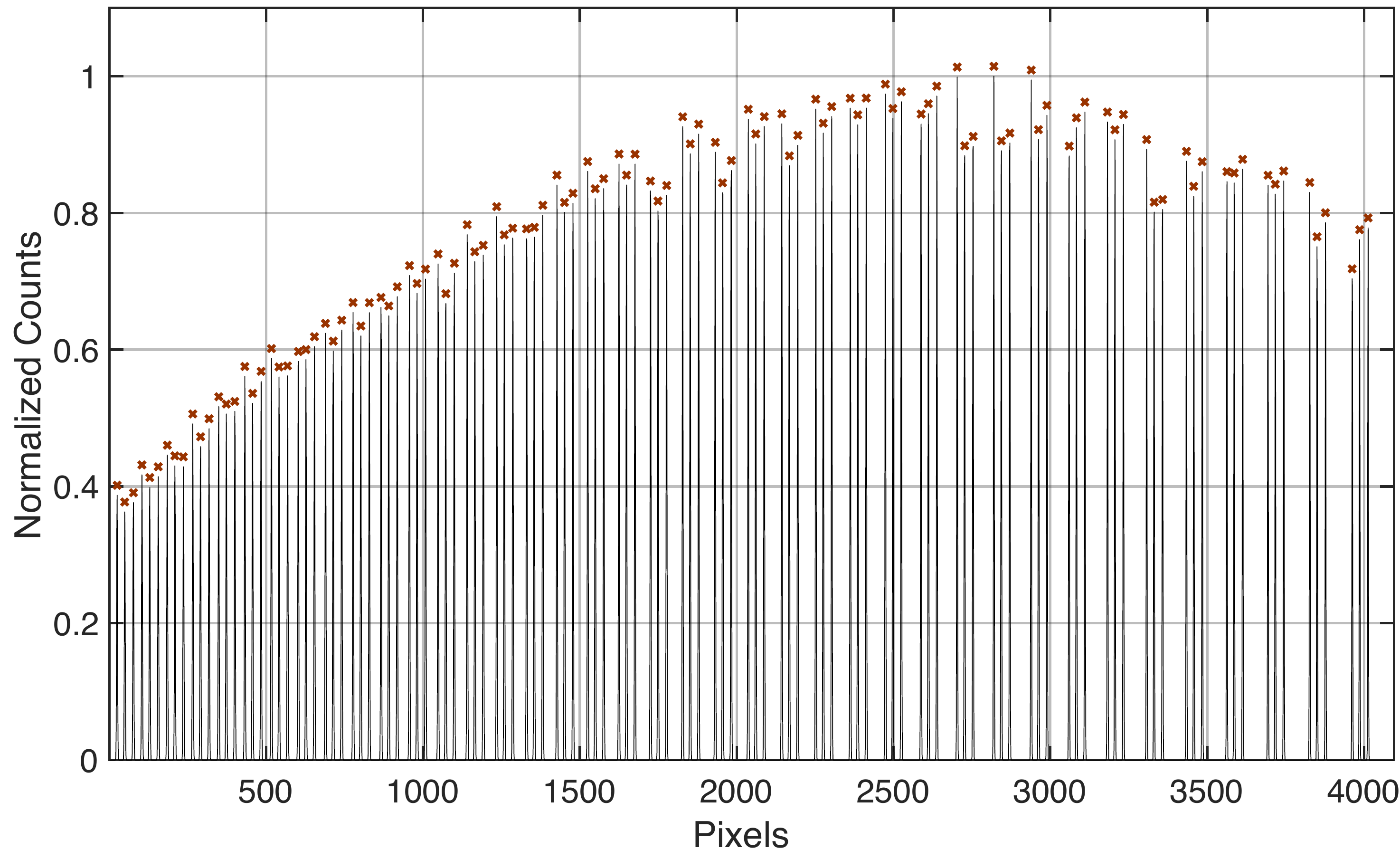}
		\label{fig:1}
	\end{minipage}%
	\begin{minipage}[t][][b]{.5\textwidth}
		\flushright
		\includegraphics[height=2.55in]{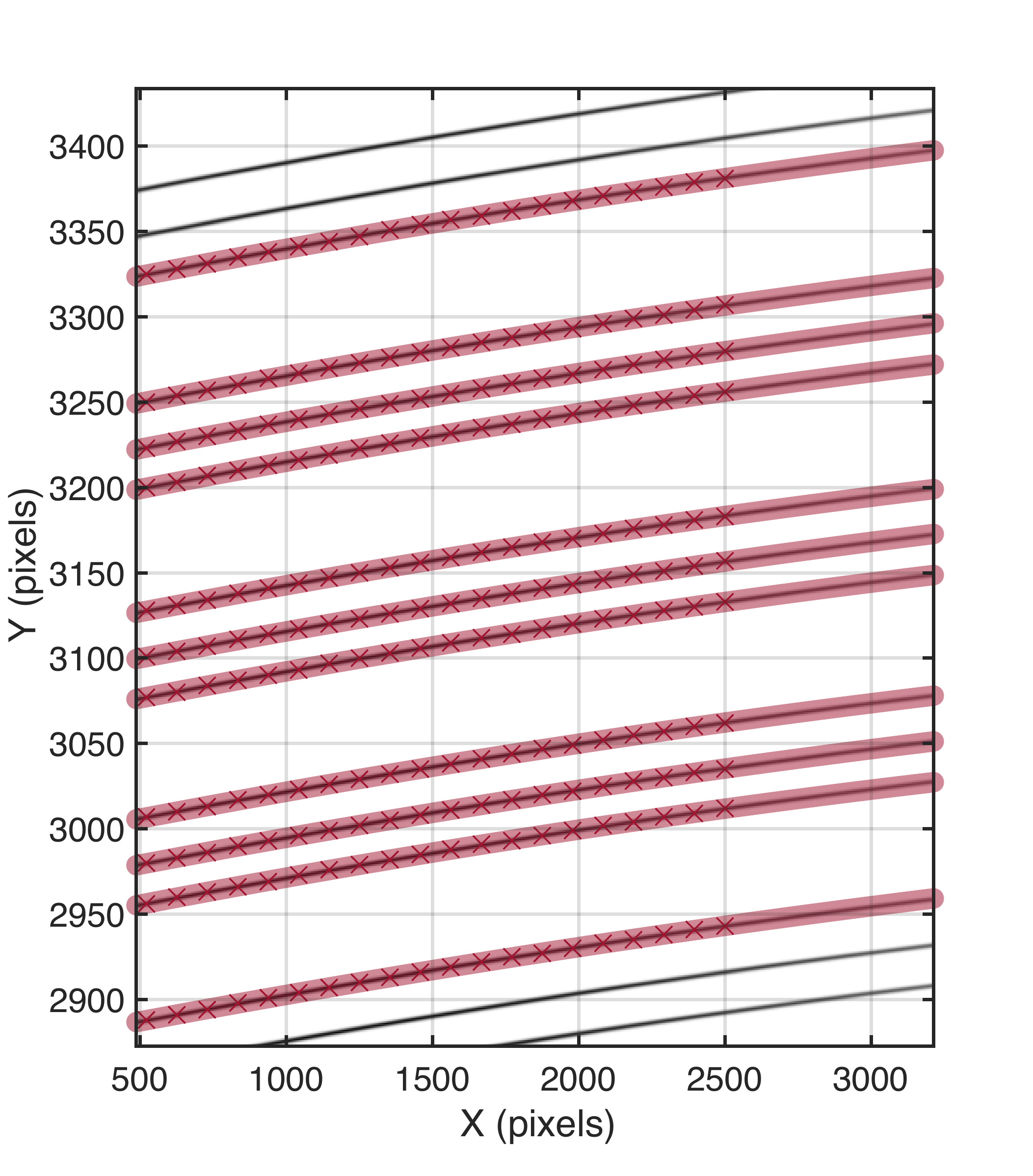}
	\end{minipage}
	\caption{\textbf{\emph{Left}:} Gaussian-smoothed, vertical slice of 10 median-combined columns showing all of iLocater's 36 spectral orders duplicated 3 times for each entrance fiber. \textbf{\emph{Right}:} Sub-frame image showing the trace algorithm and order identification algorithm working on a simulated flat field image. Color scale has been inverted to more clearly show the orders (black) and empty detector space (white).}
	\label{fig:OI}
\end{figure*}

Before a spectral order can be recovered, it must first be identified using a continuum source (super-continuum source or tungsten) to illuminate the fiber orders. Numerous vertical slices, each a few pixels wide, are taken across the detector, median combined and smoothed using a Gaussian filter. Each of these is fit using a sum of Guassians model with the number of Gaussian functions equal to the number of orders present in the slice. The resulting centroids are paired with the corresponding horizontal pixel coordinate and fit using a low order polynomial across the dispersion direction. The polynomial coefficients are saved and used for subsequent science and wavelength calibration order extractions.        
Figure~\ref{fig:OI}~(\textbf{\emph{Left}}) shows a small portion of traced orders from an image generated using the spectrograph simulator and the order identification algorithm. 

\subsubsection{Order extraction}
\label{sec:extract}
The data reduction pipeline currently has two options for performing spectral extraction: (1) a simple, sum-over-columns algorithm uses a vertical pixel window to sum all of the signal contained within, and (2) a more popular algorithm used by RV instruments, optimal extraction \citep{Horne_86,Marsh_89,Piskunov_02}, which scales 1D cross-sectional profiles to the imaged spectrum where the scaling factor is based on a flux estimate. Most algorithms will also attempt to model and reconstruct the spatial profile/slit function with a polynomial/Gaussian. 

There are also a few new algorithms being developed by other RV groups that could be added to iLocater's spectral extraction pipeline in the future. First, a ``flat-relative" optimal extraction algorithm developed by M. Zechmeister \citep{Zechmeister_14} offers improved speed and efficiency compared to classical optimal extraction. For stabilized spectrographs like iLocater, order profiles and positions are, for the most part, object- and time-independent, which means the spatial profile does not necessarily need to be modeled empirically. A high S/N master flat is used as an extraction mask where the extracted spectrum is scaled relative to the cross-section of the flat. Essentially, the extracted spectrum is measured relative to the flat spectrum and because the flat field contains the same spectral signature, e.g. spatial profile, pixel-pixel variations, and blaze function, these are all automatically incorporated into the extraction routine depending on how static they are are in practice.  

\cite{Bolton_10} outlines the ``perfect" extraction of one-dimensional spectra from two dimensional digital images of optical fiber spectrographs, based on accurate 2D forward modeling of the raw pixel data. This new technique promises statistically independent extracted samples of the 1D spectrum as well as no degradation of the 2D spectrograph resolution.However, this method requires very large matrix inversions. Another, more practical approach is offered in \cite{Kos_2018}, where photonic combs are used to precisely map aberrations. Forward-modeling convolves this map with template spectra and attempts to reproduce the observed image. Results show this reconstruction method simplifies a number of reduction steps and reliably extracts spectra with 2-3 times nominal resolution. New methods such as these represent future spectral extraction improvements that are possible for stabilized, fiber-fed instruments like iLocater.

\begin{figure*}
		\includegraphics[width=\textwidth]{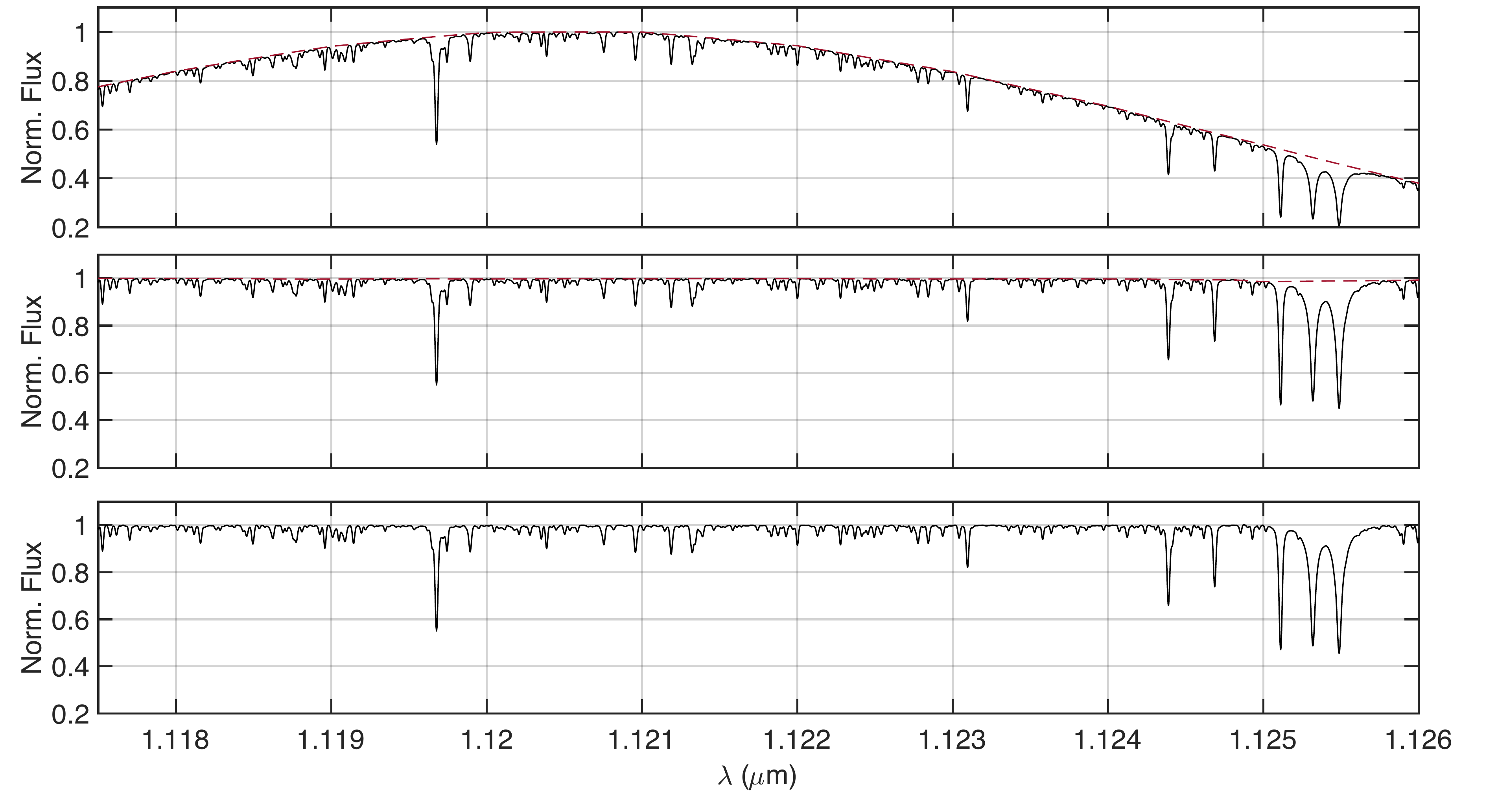} 
		\caption{\emph{\textbf{Top panel:}} Extracted spectrum of an MOV star for a particular echelle order. \emph{\textbf{Central panel:}} Same order divided by the blaze function derived from a normalized flat field image. \emph{\textbf{Bottom panel:}} Continuum normalized spectrum by fitting the deblazed continuum with a polynomial.}
		\label{fig:blaze}
\end{figure*}

\subsection{Wavelength calibration}
\label{sec:wavecal}
The wavelength calibration sub-pipeline handles tracking long-term wavelength drifts of the instrument and calculating the nightly dispersion solution. The calibration sources available to iLocater are U-Ne emission lamps and a laser-locked Fabry-P\'erot etalon. 

\subsubsection{Dispersion solution}
\label{dispersion solution}
The dispersion solution will provide science traces with a wavelength solution. To achieve this, iLocater will use a combination of the U-Ne emission lamp and a stabilized Fabry-P\'erot etalon. The Fabry-P\'erot module provides outstanding line density and stability, while the U-Ne emission lamp aids in practical identification of the etalon lines through an, absolute wavelength solution that is performed initially and periodically, as necessary. The more numerous and intrinsically stable etalon lines will be used to improve on the dispersion solution, providing a very precise wavelength calibrated spectrum, suitable for precision RV measurements. 

The wavelength calibration procedure has been adapted for iLocater, but in general follows the procedure provided in \S2.4 of \cite{brahm_16}. First, the U-Ne calibration source is fed through all three of iLocater's spectrograph entrance fibers. This illuminates every possible trace with U-Ne light, resulting in a ``full U-Ne'' exposure. Then, the etalon is used to illuminate all three fibers in the same way. The spectral extraction for both of these frames follows the same procedure described \S\ref{sec:extract}. After image processing and spectral extraction, U-Ne lines are fit using Gaussian functions and centers recorded. For most of the reduction and extraction process, spectral orders are referred to by their relative order numbers (e.g. 1-36). However, to enable a 2D wavelength solution $\lambda(x,m)$,  the actual spectral orders, m, are needed. This requires that the offset, m$_0$, is found that satisfies: 
\begin{equation}
 m_0 + i = m_i, 
\end{equation}
where $i$ represents the spectral orders numbered on the detector (e.g. $i$ = 1-36, m$_0$ = 116, $m_i$=117-152). This is done by minimizing the slope in: 
\begin{equation}
y(i)=(m_0 +i)\lambda_{\mu_i},
\end{equation}

where $\lambda \mu_i$ is the mean wavelength of the ith order \citep{brahm_16}. A 2D wavelength is then calculated using an expansion of the grating equation using Chebyshev polynomials \citep{baranne_96}. The 2D wavelength solution takes the form:
\begin{equation}
\label{eqn:2D}
\lambda_{2D}(x,m) =\frac{1}{m}\sum\limits_{i=1}^{n_m} \sum\limits_{j=1}^{n_x} c_i(m) c_j(x)\chi_{ij}
\end{equation}
where $x$ is pixel location, $m$ indicates spectral order number,  $c_i$ indicates the Chebyshev polynomial at order $i$, $n_m$ and $n_{x}$ are the degrees of the polynomial, and $\chi_{ij}$ is coefficient matrix of the wavelength solution.

\begin{figure*}[ht]
\includegraphics[width=\textwidth]{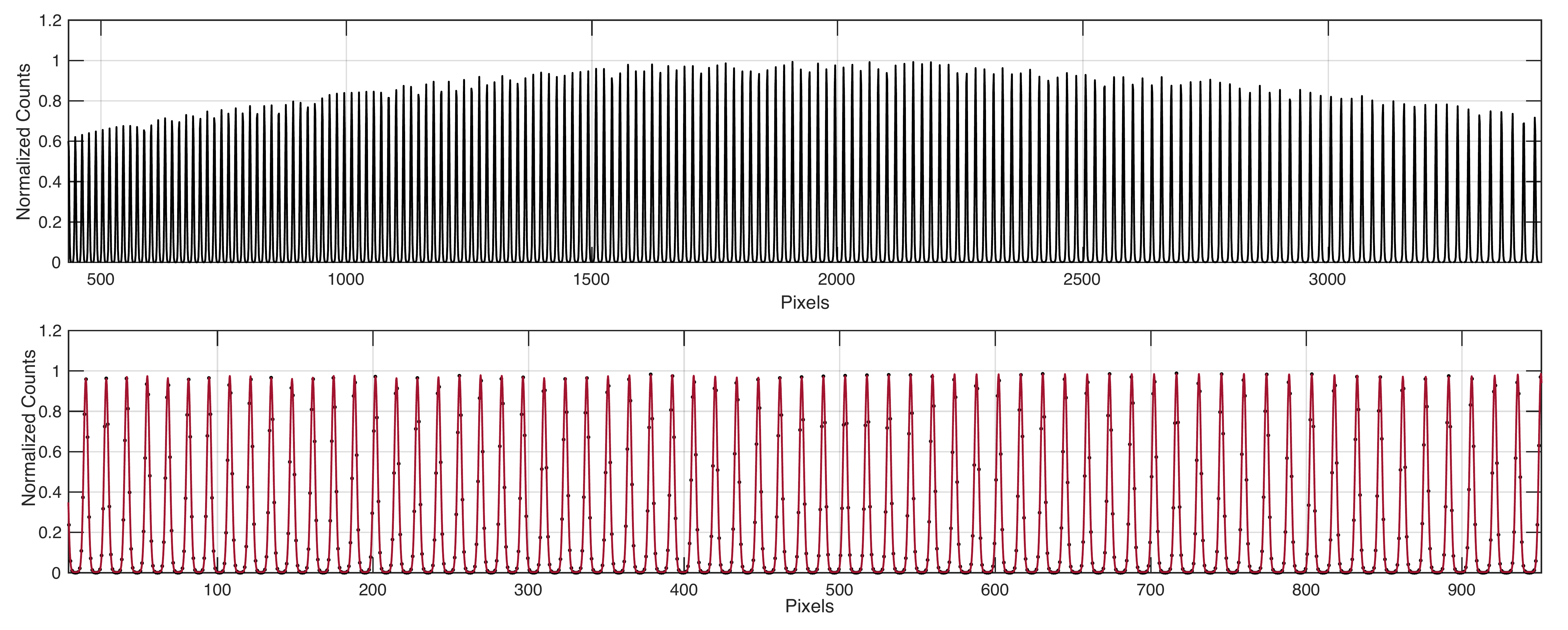} 
\caption{\emph{\textbf{Top:}} Extracted, noise-free simulation of a 10 GHz etalon, spectral order 152. \emph{\textbf{Bottom:}} Blaze corrected, normalized, sub-frame of the same spectral order demonstrating an automated, simultaneous 233 Gaussian model fit to derive pixel centers of each etalon line. The x-axis shows pixel numbers along the dispersion direction.}
\label{fig:etalon}
\end{figure*}x

\begin{figure*}[p]
		\includegraphics[width=\textwidth]{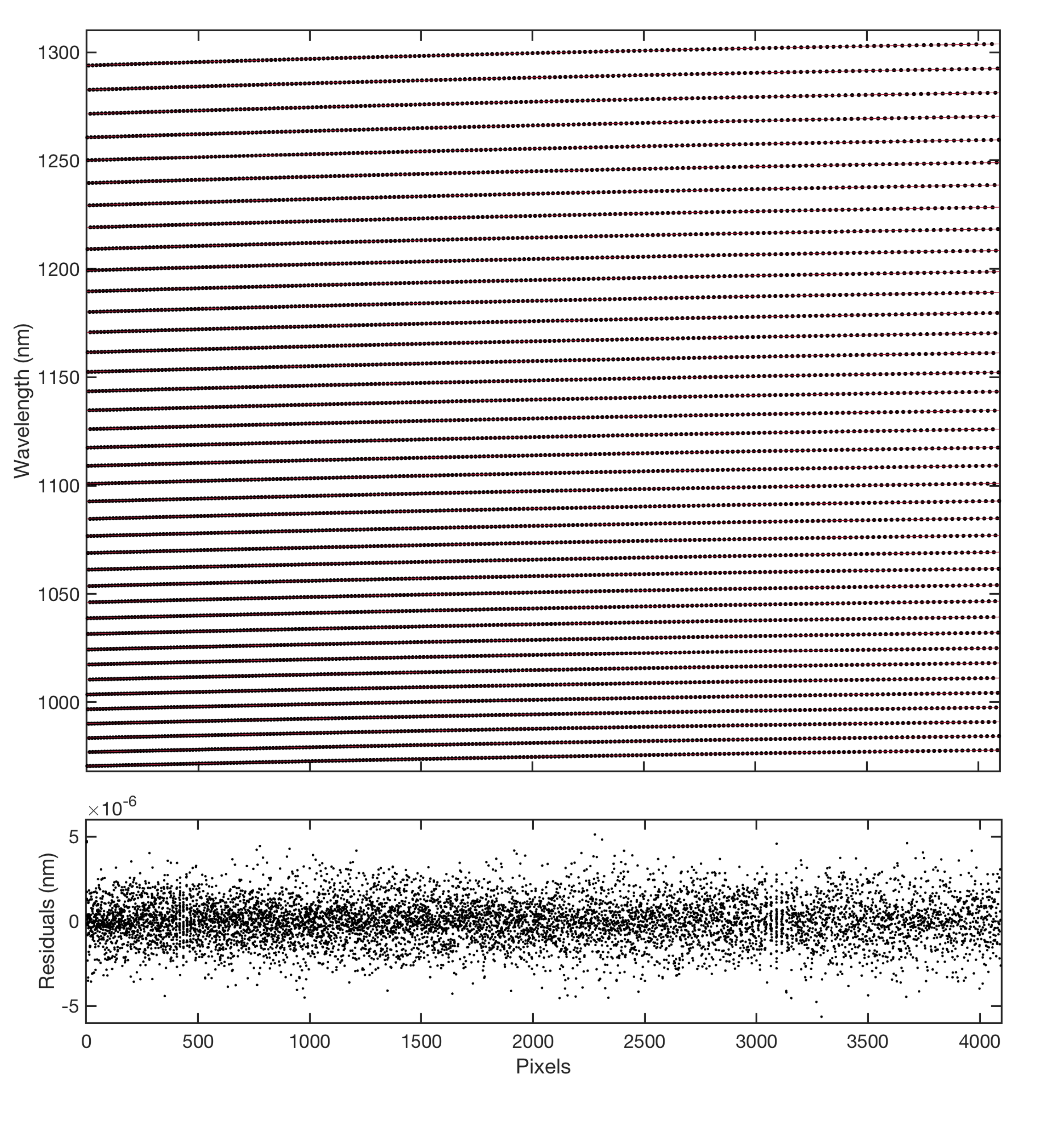} 
		\caption{2D wavelength solution derived from a simulated etalon frame. Points represent etalon peaks in wavelength and x-pixel direction while the black line denotes the 2D wavelength solution. Residuals are shown below and have an rms of $1\times10^{-6}$ nm.} 

		\label{fig:2d}
\end{figure*}

Next, the extracted etalon spectral orders are fit using a multi-Gaussian model, order by order, to determine each etalon peak in pixel space. The specific model used is a sum of Gaussians:

\begin{equation}
	\sum\limits_{i=1}^{N_{Gauss}} a_i \cdot e^{\frac{{-(x-b_i)^2}}{2\cdot c_i^2}} + d_{off} + x\cdot e_{slope}
\end{equation}
where $N_{Gauss}$ indicates the number of Gaussian functions chosen to fit an order, $a_i$ is the amplitude free parameter, $b_i$ is the centering parameter, $c_i$ is the width parameter, $d_{off}$ is a single global offset parameter, and $e_{slope}$ is a global linear slope parameter designed to account any residual continuum during the fitting process. Figure~\ref{fig:etalon} shows a single order of an extracted etalon spectrum with an overlapping best-fit model in red. As the etalon FSR results in an extremely regular and close spacing of emission peaks, differentiating individual peaks can be difficult. Using the U-Ne wavelength solution, an existing line list of etalon peak wavelength centers, or a combination of both, a wavelength is assigned to each etalon peak. Now the etalon can be used to derive a 2D wavelength solution in the same form as Equation~\ref{eqn:2D}, but with a precision far exceeding the limitations of U-Ne (Figure~\ref{fig:2d}). U-Ne frames used to create an initial wavelength calibration are valid as long as the instrument remains stable enough that etalon lines will not spatially drift so much they would be mistaken for an adjacent peak. As iLocater is intended to be very stable over long periods of time, U-Ne calibration will rarely be used in comparison to the etalon.   

\subsubsection{Instrument drift}

iLocater has three simultaneous traces for each spectral order, fed by two fibers, one connected to each telescope, and one simultaneous reference fiber that is set between the two science fibers. During science exposures, and between bracketed calibration frames, this reference fiber is used to correct for instrument drifts on timescales comparable to the integration time. Drifts are calculated using Equation~\ref{eqn:2D} but holding the  $\chi_{ij}$ coefficients constant and introducing a new free parameter, $\delta v_{drift}$, the velocity drift, to Equation~\ref{eqn:2D}:

\begin{equation}
\lambda_{2D}(x,m) =\frac{1}{m}\bigg(1+\frac{\delta v_{inst}}{c}\bigg)\sum\limits_{i=1}^{n_m} \sum\limits_{j=1}^{n_x} c_i(m) c_j(x)\chi_{ij}.
\end{equation}

\subsection{Post-processing}

The post-processing sub-pipeline will produce the finalized spectral data product. The remaining steps included in post-processing are: blaze function removal, continuum normalization, telluric corrections, and barycentric correction. 

\subsubsection{Telluric correction}

Telluric removal in the infrared requires careful consideration and is extremely difficult to achieve a correction residual of less than 1 m/s \citep{bean_10}. Currently, most groups attempt to model telluric lines with synthetic models using a comprehensive line list and radiative transfer with accurate atmospheric models. iLocater's telluric removal strategy will follow a similar process to TAPAS \citep{Bertaux_14}, TelFit \citep{Gullikson_14}, Molecfit \citep{Smette_15}, and TERRASPEC \citep{Bender_12}, using the line-by-line radiative transfer model \citep{Clough_05}; and the High Resolution Transmission (HITRAN) line database \citep{Rothman_13} to generate and subtract a representative synthetic telluric spectrum. This method is typically more accurate than empirical correction \citep{Gullikson_14, Smette_15}, achieving line removal precision of 2-5\%. Poorly fitted lines will be masked out following the techniques in \cite{bean_10, Seifahrt_10, Blake_10}. 

None of the current techniques have yet reached the necessary precision for sub meter-per-second measurements in the NIR. The reasons for this are primarily missing lines in the HITRAN database, uncertainties or errors in attributes such as the line position, strength, or shape, limitations in current modeling codes for deriving correct line profiles (i.e., velocity dependence, line mixing effects), insufficient knowledge of real time atmospheric conditions (e.g. water column density variations), and wind-induced line shifts \citep{fischer_16}. 

\subsubsection{Sky-background subtraction}

iLocater is using a $5.8~\mu m$ diameter single-mode fiber that has a very small on-sky angle of $1.4\times10^{-3}~\mathrm{arcsec}^2$, compared to a typical large diameter multi-mode fiber angle of $1~\mathrm{arcsec}^2$. Because of this, there is significant natural suppression of sky-background emission, making dedicated sky measurements and removal procedures largely unnecessary. However, in the case that iLocater requires simultaneous sky measurements, one of the science fibers could be used to sample sky background while the other collects starlight using the LBT's differential pointing capabilities. 

\subsubsection{Blaze Removal}
\label{blazeremoval}
The efficiency function of each order in an echelle spectrograph is normally dominated by a characteristic slope known as the blaze function which must be corrected prior to RV computation. A measurement of the blaze function can be made by injecting a flat field source into the fibers. The effect of the blaze function can be mitigated by normalizing and dividing the science spectrum by the fiber flats (Figure~\ref{fig:blaze}). However, this will not completely correct the spectrum because of the difference in continuum between the observed star and flat field source. Varying polarization states in time resulting from intrinsic single-mode fiber properties will also cause a slightly different blaze measurement each time it is measured, emphasizing the need for further continuum normalization. 

\subsubsection{Continuum normalization}

The remaining residual continuum left in the spectrum after de-blazing must be removed as it can result in spurious Doppler measurements. The continuum is sampled using a moving box that rejects absorption lines. The bottom panel of Figure~\ref{fig:blaze} shows very minor residual slopes being removed. Continuum normalization should be applied very cautiously as in some cases the correction induces its own residual tilt if the continuum is improperly sampled.   

\subsection{Barycentric correction}

When using the RV method to search for planets orbiting nearby stars, the dominant signal is due to the Earth's motion about the solar system barycenter. Barycentric correction consists of computing the observatory velocity with respect to the solar system barycenter, projected in the direction of the observed star. The two principal velocities to be computed are the movement of the Earth around the barycenter and the Earth's rotation at the geographical coordinates of the observatory. We implement the \cite{wright_14} correction which uses full, relativistic, calculations precise to 1 cm/s. 

\subsubsection{Exposure meter}
Barycentric corrections can only be calculated for a single instant in time, however, iLocater's exposure times are expected to last 30 minutes or longer. Additionally, the barycentric correction does not scale linearly with time so a flux-weighted average needs to be calculated throughout the observation (approximately every 1 minute to achieve 1 cm/s correction error). To calculate this flux-weighted average, iLocater will use readouts from a femptowatt photoreciever, located just before the entrance fiber in the spectrograph. For a ground-based observation, atmospheric extinction will introduce a wavelength dependence in the transmittance of photons to the instrument, possibly requiring a wavelength dependent barycentric correction. iLocater's H4RG detector is capable of non-destructive readouts which could potentially be used as a chromatic exposure meter by calculating a different barycentric correction for each spectral order.   

\subsection{Radial velocity determination}

\subsubsection{Building a binary mask}
\label{maskbuilding}
\begin{figure}
	\includegraphics[width=\columnwidth]{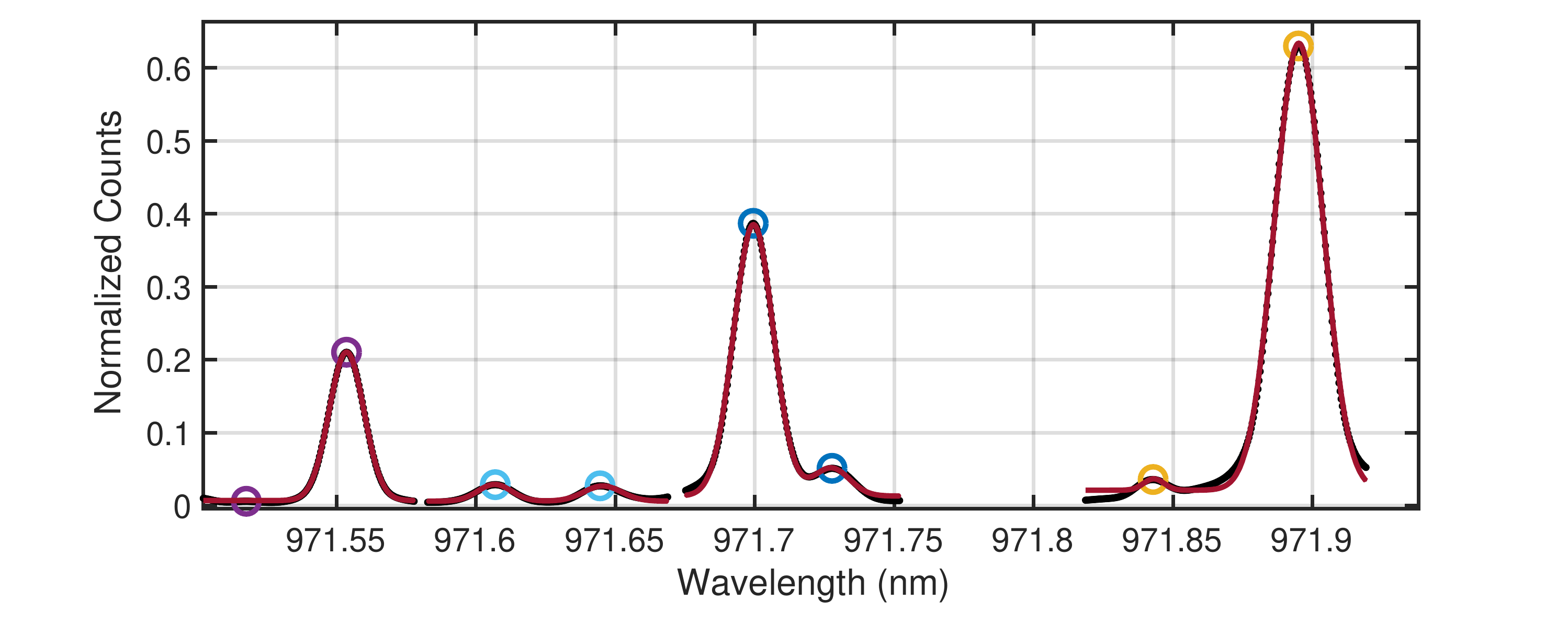}
	\caption{Small subsection of the mask building process showing an inverted synthetic stellar spectrum along with all identified peaks(circles) and multi-Gaussian fits(red line).} 
	\label{fig:maskfit} 
\end{figure} 
A binary mask used in cross-correlation serves the purpose of only considering the RV information contained in certain selected stellar absorption lines. This helps avoid blended lines, lines with minimal depth compared to the continuum, and lines with other unsuitable characteristics. In general, the mask should include as many `clean' lines as possible to maximize the RV signal. Additionally, lines should be weighted by their depth relative to the normalized continuum. 

We have written a \textsc{Matlab} function to scan across a synthetic stellar spectrum, given a spectral type, rotational velocity, and bandpass, and builds a line list for masked cross-correlation. First, the synthetic spectrum is normalized, its sign inverted and brought to a range of [0 1]. The spectrum is divided into 0.1~nm pieces and a peak-finding function computes all observable peaks. A small interval is chosen around the first and last peak in the piece and is fit with a multi-Gaussian function where the number of Gaussians is equal to the number of peaks found. Each Gaussian has 3 parameters describing its amplitude, center, and width and 2 parameters for a constant offset and linear slope to remove any residual local trends in the inverted continuum. An example of this process is shown in Figure~\ref{fig:maskfit}.  

\begin{figure}
 \includegraphics[width=\columnwidth]{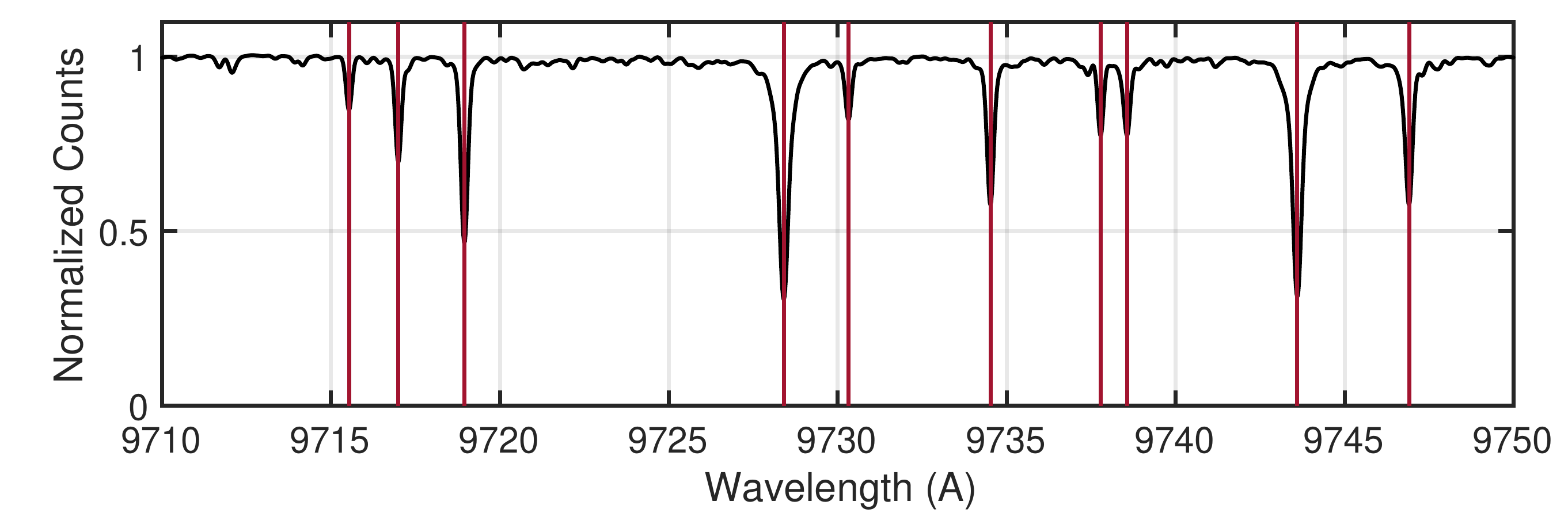}
  \caption{Masked stellar lines for a portion of order 152 overlaid on an M0V synthetic spectrum.} 
  \label{fig:mask} 
 \end{figure}

The line list is formed by appending the Gaussian center parameters as well as the ratio of the amplitude to the offset value. In order to avoid blended lines, which are present in many spectral chunks, the software identifies pairs of lines whose separation is less than a critical value. In this situation, only the line with largest amplitude is left in the mask list. To further clean the line list from unsuitable lines, only lines with a value of the FWHM roughly in agreement with that expected for the rotational velocity of the star are kept. Figure~\ref{fig:mask} shows a portion of a simulated order (order 152) of a normalized M0V spectrum with the final line list overlaid in red.   

\subsubsection{Cross correlation function} 
\label{CCF}
The RV information of an observed star is contained within the wavelength positions of its spectral absorption lines, specifically in the line ``wings'', i.e. where the derivative is maximized. An efficient way to measure the variation of spectral absorption lines in time is using the cross-correlation function (CCF), with a binary mask taking values equal to 1 in the regions where a typical stellar spectra contains narrow absorption lines and equal to 0 elsewhere. 
For the RV extraction sub-pipeline, we have written our own version of the standard masked CCF technique \cite{baranne_96, Pepe_02}. This technique is currently used by the HARPS team and has performed very well in precision RV applications \citep{bonfils_13}. The CCF is given by: 
\begin{equation}
	CCF(v) = \frac{\int_{\lambda_i}^{\lambda_f} w(\lambda')m(\lambda')s(\lambda) d\lambda}{\sqrt{\int_{\lambda_i}^{\lambda_f}m(\lambda')^2}d\lambda} , 
\end{equation}
where $\lambda_i$ and $\lambda_f$ are the initial and final wavelengths of an order, $s(\lambda)$ is the spectrum, $w(\lambda')$ is a list of weights stored in the binary mask, $m(\lambda')$, and $\lambda'$ is the Doppler shifted wavelength \citep{brahm_16}. 

\section{Verification of Simulator \& Pipeline Performance}

\subsection{Numerical simulation errors} 

\begin{figure}
	\includegraphics[width=\columnwidth]{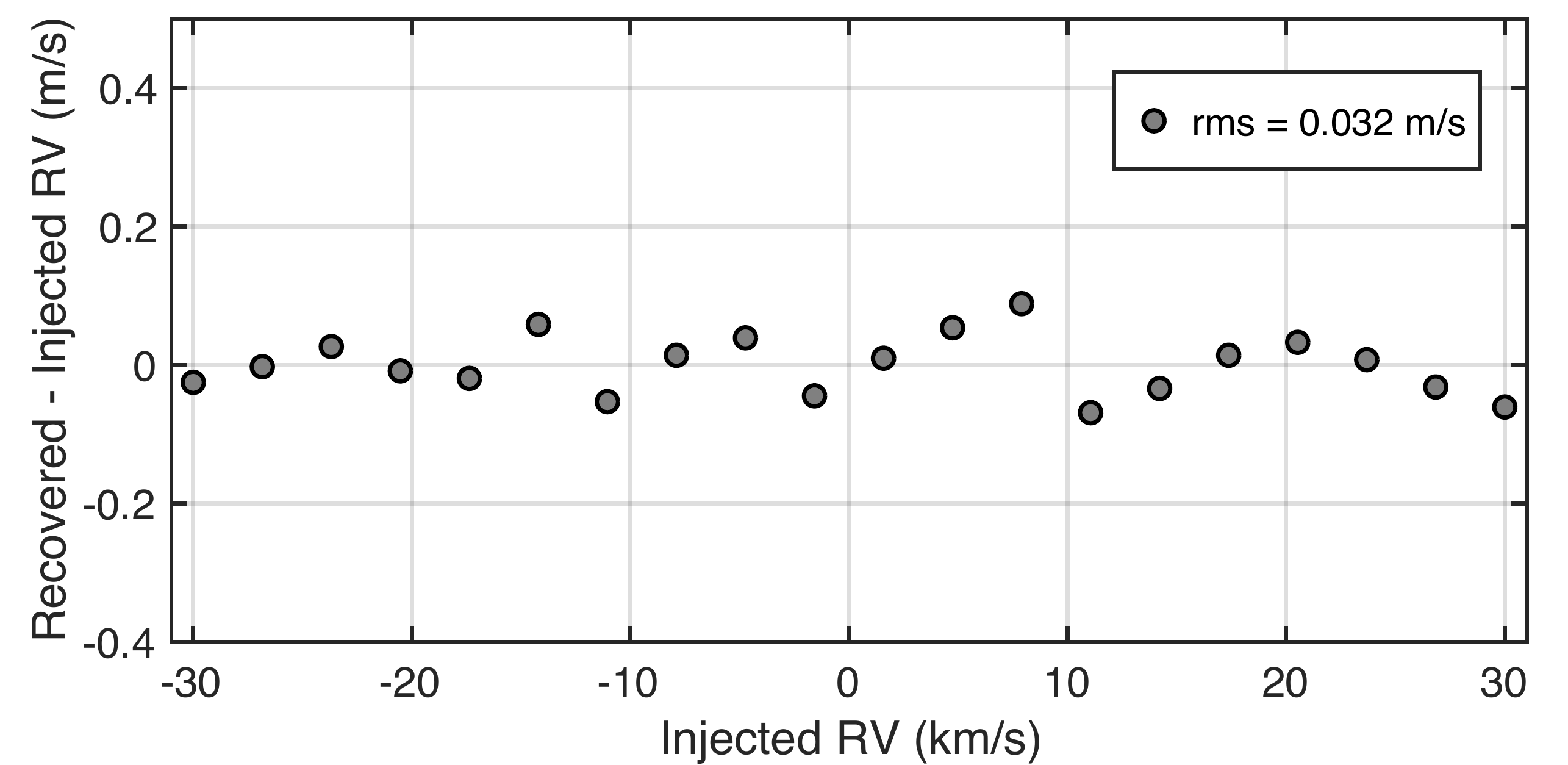}
	\caption{RV residuals from simulated noiseless data frames where a synthetic stellar spectrum (M0V) was shifted by precise Doppler velocities between -30 km/s to +30 km/s. Velocities were then recovered using masked cross correlation (\S\ref{CCF}). These residuals represent the inherent algorithmic noise injected through data reduction processing.}  
	\label{fig:residualRV} 
\end{figure}

Measuring Doppler shifts on the order of 10 cm/s requires a very precise software pipeline. This also requires that the simulation code is not introducing algorithmic or numerical errors at a level that would overwhelm this measurement. To quantify the simulator's numerical errors in RV values, we simulated data frames of an M0 star. These frames contain no physical noise, no throughput modifications, or any modifying effects in order to isolate computational errors from simulated instrument systematic errors. 20 total frames were simulated, varying only injected RV, between -30 km/s and 30 km/s, chosen to reflect the typical magnitude of RV fluctuations under barycentric motion. The data reduction pipeline was then used to extract each order, apply an ideal wavelength solution, and compute the RV using masked cross correlation, which also serves as a test of the pipeline's ability to recover RV's under noiseless conditions. The residual error measured between the pipeline's recovered RV and the injected RV into the simulator are shown in Figure~\ref{fig:residualRV}. The residual scatter shows structural errors do not exceed $\mathrm{10~cms^{-1}}$, with an $\mathrm{rms=0.03~cm/s}$. From this, we conclude numerical simulation errors are sufficiently suppressed for the purposes of probing design choices and developing robust extraction/analysis software. Note that this test was performed under a simulation scale factor of 1, described in Table~\ref{tab:inputs}. Numerical errors can be further reduced by increasing the scale factor but at the cost of computation time, where the time to complete a simulated frame approximately increases with the square of the scale factor.

\subsection{Masked CCF performance}
\label{maskperformance}
\begin{figure}
	\includegraphics[width=\columnwidth]{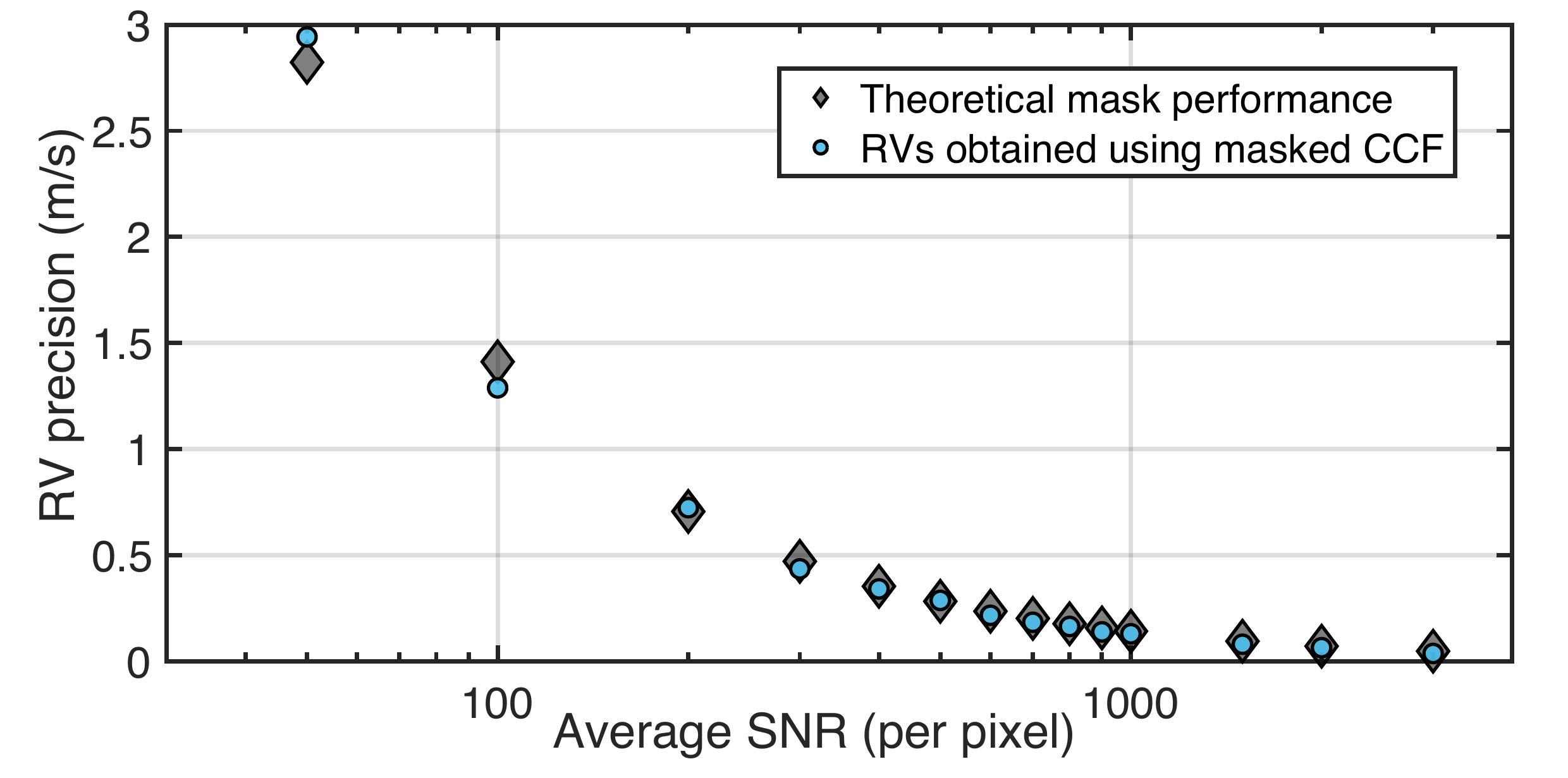}
	\caption{The RMS scatter in RV residuals for a set of simulation runs with varying levels of photon noise, as a function of collapsed SNR per pixel. The implementation of a masked CCF RV measurement routine approximates the expected ideal performance for the mask used. This set of results focus on mask performance and do not include sources of error other than photon noise.}  
	\label{fig:maskperformance} 
\end{figure}

Another test of the simulator and pipeline performance is to verify the quality of stellar masks derived in \S\ref{maskbuilding} using photon noise to investigate various signal-to-noise levels. This experiment is conducted by varying the average SNR, per collapsed pixel, in simulated frames. Here, collapsed pixel is referring to the conversion of 2D extracted spectral orders to 1D collapsed orders by recombining signal in the cross-dispersion direction. Masked CCF performance is measured against the Doppler information content contained in iLocater's spectral orders, measured using the formalism in \cite{butler_96}: 

\begin{equation}
\sigma_{\rm ph}=\frac{1}{\sqrt{\sum \left(\frac{dI/dV}{\epsilon_I} \right)^2}},
\end{equation}
\label{eq:dopplerquality}
where $dI/dV$ represents the slope of the measured stellar intensity as a function of wavelength (expressed in velocity units) and $\epsilon_I=\sqrt{N_{\rm ph}}/N_{\rm ph}$ is the fractional Poisson error. Figure~\ref{fig:maskperformance} shows the achievable RV precision computed at each SNR as well as the recovered RV from masked cross-correlation, measured by the RMS of 50 repetitions at each SNR. This shows the simulated photon-noise routines, as well as the derived masks are close to the expected performance of a masked CCF routine.        

\section{Applications}

\subsection{Assessing instrument performance}

\begin{figure}
\includegraphics[width=\columnwidth]{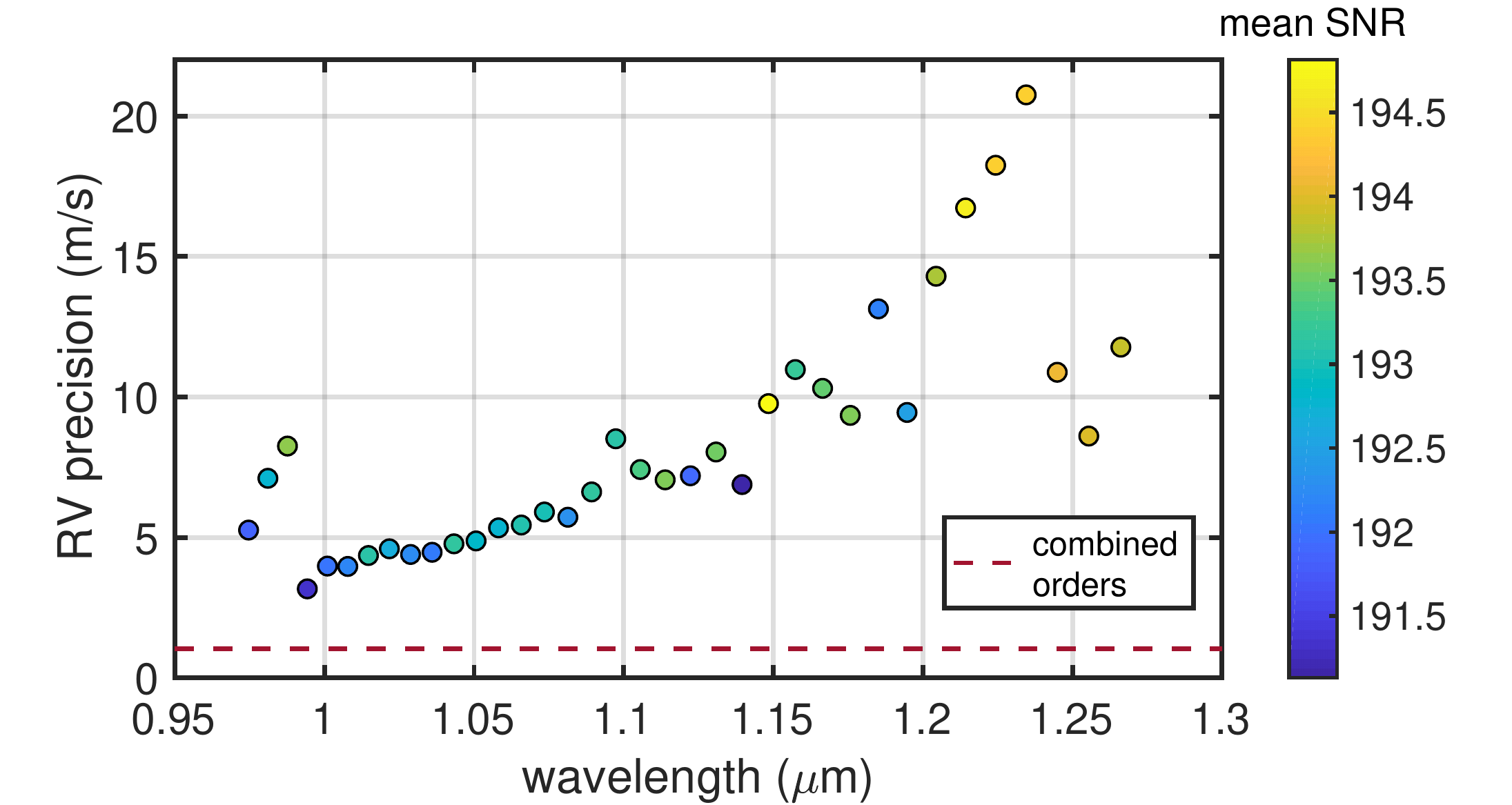}
\caption{Achievable RV precision for each of iLocater's 36 spectral orders. The target chosen was an M0V star with an average SNR between 190 and 195 across the spectral orders. Combining the RV information in every order gives a single measurement RV precision of 1.02 m/s.}
\label{fig:phnoise}
\end{figure}

One of the first and most important uses of the software tools described within is to determine achievable on-sky photon-noise limited signal-to-noise. This is accomplished by using the simulator to generate different spectral types at a range of apparent magnitudes, integrating the signal on the detector over the desired integration time, and then assessing the achievable RV precision using the formalism established in \S\ref{maskperformance}. Photon noise limited RV precision, $\sigma_{ph}$, is calculated using Equation~\ref{eq:dopplerquality}. This could be used is to assess the photon-noise limited RV precision contained in each of iLocater's spectral orders for a specific spectral type and average SNR per order. Figure~\ref{fig:phnoise} shows each of iLocater's individual spectral orders containing an M0V star with approximately constant SNR. This plot shows there is a clear trend in achievable Doppler precision, decreasing with longer wavelengths as well as the necessary average SNR in order to achieve m/s RV precision. 

\subsection{Spectral order cross-contamination}

\begin{figure*}
		\includegraphics[width=\textwidth]{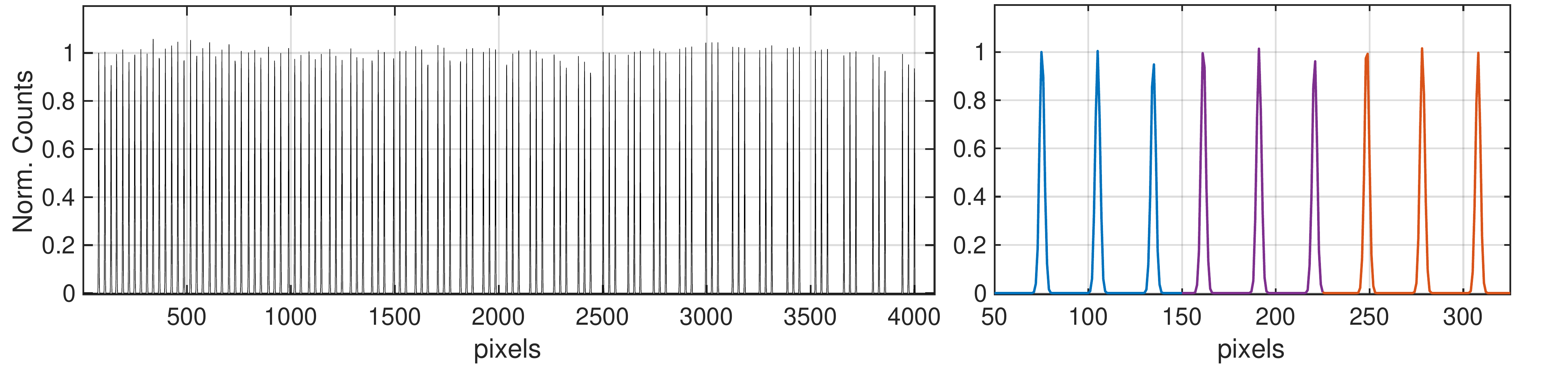}
		\caption{\emph{\textbf{Left}}. Slice of uniformly illuminated fiber traces on detector in the cross dispersion direction showing 36 spectral orders with 3 traces in each order. Note, there is no modulation in counts as throughput effects are not included here. \emph{\textbf{Right}}. Windowed plot focusing on the first three spectral orders, going left to right. Each spectral order is identifiable by a different color.}
\label{fig:contamination}
\end{figure*}

Choosing the correct cross-disperser in echelle spectrograph design is an important step to avoid adjacent orders contaminating each other while maximizing information content by optimizing the number of orders that are captured by the detector. We use the simulator to visualize the optical design and inspect the inter-order and intra-order separation. For iLocater, we set an upper limit of less than 0.1\% cross-contamination between orders. Therefore, we define the `edge' of an order to be the point at which the ratio of intensity relative to maximum is 0.1\%. As the vertical profile of each order follows a Gaussian profile, this quantity is easily computed, given a known FWHM. Adopting a pessimistic value of 5 pixels sampled in a FWHM in the cross-dispersion direction, this gives a requirement of 16 pixels between adjacent order centers. To test this, we generate a noiseless full frame detector image with a flat-field spectrum in each fiber trace. Vertical cross cuts are taken with a focus on spectral orders with the shortest wavelengths as they will experience the smallest separations. Figure~\ref{fig:contamination} shows a cross-cut of all 36 spectral orders and 3 traces within each order. The right side shows a plot of the first three spectral orders, where the closest separation exists between the third trace of order one and the first trace of order two. Fitting these with Gaussian profile and computing the separation gives a value of 27.3 pixels, well within the contamination requirement.  

\subsection{Optimizing Fabry-P\'erot finesse and free spectral range}
\label{optimize etalon}

We derive optimal finesse ($\mathcal{F}$) and free spectral range ($FSR$) values for our Fabry-P\'erot etalon  and verify these calculations using visuals created with the simulator. First, we define three straightforward requirements for the etalon:

\begin{enumerate}
	\item The etalon needs to span the primary science bandpass of iLocater (0.97-1.27 $\mu$m) to provide satisfactory wavelength calibration information in every spectral order.
	\item To maximize information content in the calibration source and simplify the peak-fitting process, the etalon modes should not be resolved by the spectrograph.
	\item In order to optimize the number of peaks in each order but avoid overlapping, we require that the minimum line separation (peak-to-peak) be at least 3 PSF FHWM of the spectrograph PSF. 
\end{enumerate}
From these requirements, we can derive the spectral characteristics of an optimal Fabry-P\'erot etalon cavity, specifically the finesse, which sets the etalon peak width, and free spectral range which defines peak-to-peak separation. 

\subsubsection{Finesse}
Following the formalism presented in \S3.1 of \cite{Cersullo_17} and using iLocater's bandpass and an under-sampling factor of 5 instead of 3, we obtain the following condition for the finesse:
\begin{equation}
\mathcal{F} = 15 \cdot \frac{ \lambda_R}{\lambda_B}. 
\end{equation} 
Using iLocater's wavelength bounds ($\lambda_B = 0.97 \; \mu m$ and $\lambda_R = 1.27 \; \mu m$) we specify a minimum finesse of 20. Due to small batch-to-batch reflectivity ripple in the custom-manufactured etalon mirror surfaces, the finesse of the final system is being targeted at $\sim$40, with an expected margin of 20-70. This range fulfills our requirement.

\begin{figure*}[ht]
\begin{center}
		\includegraphics[width=0.8\textwidth]{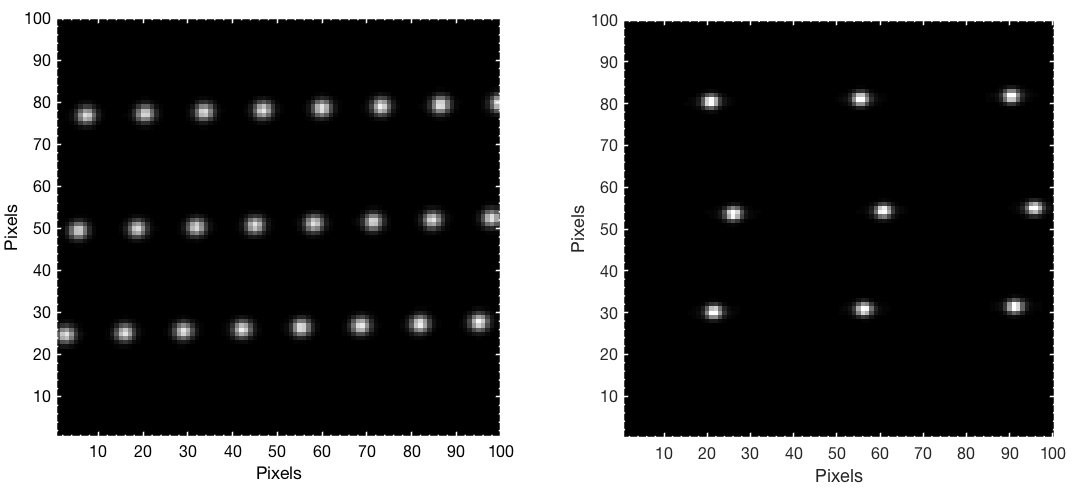}
		\caption{{\emph{\textbf{Left}}}. Simulated detector frame centered at the shortest wavelengths of iLocater's bandpass. This region corresponds to the closest etalon spacing and visually confirms sufficient separation according to requirement (2). \emph{\textbf{Right}}. Simulated detector frame centered at the longest wavelengths. Etalon peaks still appear unresolved, adopting the shape of the instrument PSF. 
}
\label{fig:etalonspacing}
\end{center}
\end{figure*} 

\subsubsection{Free spectral range}
Condition (3) requires there to be a minimum separation equivalent to 3 PSF FHWMs at the bluest wavelength. For this part of the calculation we will adopt a more conservative definition of PSF width, moving from FWHM to 1/e$^2$ ($\approx$ 13.5\%). Using the following relation to compute Gaussian width at intensity, I, for a known $\sigma$,   
\begin{equation}
Gauss~width(I) = 2\cdot \sigma \cdot \sqrt{2\cdot \log\Big(I^{-1}\Big)},
\end{equation}
with I = 1/e$^2$ gives an equivalent separation of 5.1 FWHMs, slightly modifying the value given in condition (3). Inserting this value in Equation 7 of \cite{Cersullo_17} gives:
\begin{equation}
FSR_B \geq 5.1 \cdot \frac{\lambda_b}{R}
\label{eq:FSRB}
\end{equation}
where $\lambda_B$ is the bluest spectral bandpass wavelength and $R$ is the spectrograph resolution. Computing Equation~\ref{eq:FSRB} with $R$ = 150,000 and $\lambda_B = 970$ nm results in a $FSR = 10.5$ GHz. 

To visualize the results derived above, we generate a synthetic etalon spectrum with an $\mathcal{F}$ = 40 and $FSR$ = 10 GHz and pass it through our spectrograph simulation code. Figure \ref{fig:etalonspacing} shows two 100$\times$100 pixel windows taken from the full simulated detector frame. The left frame is centered on the shortest wavelengths of the bluest spectral order, showing that even at the minimum etalon peak separation, each peak is well separated from adjacent peaks. The right frame centers on the longest wavelengths of the reddest order, showing the maximum separation of each order.  

\subsection{H4RG detector noise}
iLocater is implementing an H4RG-10 NIR detector. Although it is relatively untested in Doppler spectrographs compared to the H2RG, it offers the necessary array and pixel size to accommodate iLocater's spectral resolution, bandpass and pixel sampling needs. An independent study has been undertaken to focus on the suitability of H4RG detectors used in precision RV work by translating their noise characteristics into RV errors (Bechter E. et al. 2019 (a), \textit{in prep.}). This work makes use of the spectrograph simulator, HxRG noise generator, and data reduction pipeline to characterize detector noise and unique NIR detector effects and translate them directly into RV errors in addition to investigating potential mitigation strategies.



\subsection{RV error budget}
Modern Doppler spectrographs are pursuing extraordinary precision in single RV measurements, seeking to reduce instrument systematic errors and maximize resolving power to disentangle stellar activity signals from observed spectra \citep{Davis_17}. Many instruments are investigating new and relatively untested designs to push the current precision boundaries. To verify their fidelity, quantified RV error budgets are becoming standard practice during the spectrograph design phase, in which each design choice is assessed by its impact on RV precision \citep{halverson_16}. Using the simulation tools outlined in this paper, a comprehensive RV error budget that includes: photon noise, instrument systematic terms, software and barycentric removal residuals, and time-varying atmospheric contaminants has already been assembled for iLocater, the details of which can be found in \cite{BechterA_2018}.      
    
\subsection{Optical aberrations}

Optical aberrations in spectrographs originate from many possible sources including: manufactured optical surfaces, misalignments in the optical system, thermal effects, etc. If the aberrations vary over time, they can impart asymmetries on the spectral signal through the instrument PSF, leading to a shift in the center of the light distribution on the detector focal plane. Typically, spectrograph optical designs are optimized using spot sizes or wavefront error budgets. We use the simulator and pipeline in a sensitivity study specifically designed to investigate the impact of optical aberrations on RV measurements (Bechter E. et al. 2019 (b), \textit{in prep.}).

\section{Summary}

Astronomers are building planet-finding spectrometers that aim to measure Doppler shifts at the level of 1 m/s and below. The ultimate goal of studying the masses and orbits of terrestrial worlds in the habitable zone requires RV precisions that are an order of magnitude better than the current state of the art. This level of performance corresponds to routinely measuring translational line shifts of only several atomic radii on the detector. A number of subtle effects can impact the retrieval of these extraordinarily small RV variations. 

Data reduction pipelines play a key role in being able to reliably extract the motion of stellar absorption lines in RV time-series data. Software represents not only an error term itself, but also permits the evaluation of many other terms in the error budget through numerical simulations. Developing a data reduction pipeline early-on in the instrument design and development process is thus essential as it provides insight into the effects that limit precision.   

We have developed a comprehensive simulation code and RV data reduction pipeline for the iLocater spectrograph, an AO-fed SMF Doppler instrument being constructed for the LBT in Arizona. This paper provides a detailed description of the code structure and overview of the how the various classes interact. This infrastructure has been used to inform design decisions for the spectrograph and quantify RV error budget terms. 

Using conventional methods for cross-correlating stellar spectra, we find that iLocater's software pipeline typically results in residual RV variations of several centimeters per second. This level of precision is well below the effects introduced from astrophysical jitter or that expected from photon noise and instrument stability. Among other things, these tools have been used to study the impact of signal-to-noise (e.g. throughput budget) and to optimize and verify wavelength calibration line spacings, order spacing, masked-CCF performance, and fiber contamination effects. In the future, the pipeline will be used to further study the effects of optical aberrations, barycentric correction, telluric contamination, the use of HgCdTe devices, and characterizing absorption line asymmetries at high spectral resolution.  

\acknowledgements
We thank Ryan Terrien and the rest of the HPF and NEID teams for many useful discussions throughout the development of the simulator and data reduction pipeline. Julian St\"urmer and the MAROON-X team have also provided helpful suggestions in regards to simulating and deriving wavelength solutions from a Fabry-P\'erot etalon. We would also like to thank former students in our group Edward Kielb and David Shaw for their early efforts with the simulator project. Justin R. Crepp acknowledges support from the NASA Early Career and NSF CAREER Fellowship programs. 
\software{\textsc{Matlab} \citep{MATLAB}, nghxrg \citep{Rauscher_2015}}

\bibliography{report}   


\end{document}